\documentclass[iop]{emulateapj} 

\begin{document}

\newcommand{\cl}{{RX~J0152-13}}
\newcommand{\ms}{{MS~1054-03}}
\newcommand{\clt}{{CL~1358+62}}

\newcommand{\g}{$g_{475}$}
\newcommand{\B}{$B_{435}$}
\newcommand{\V}{$V_{606}$}
\newcommand{\ra}{$r_{625}$}
\newcommand{\ia}{$i_{775}$}
\newcommand{\I}{$I_{814}$}
\newcommand{\z}{$z_{850}$}

\newcommand{\ntot}{62}
\newcommand{\nhst}{59}
\newcommand{\nmag}{44}
\newcommand{\ndispm}{31}
\newcommand{\ndisp}{36}
\newcommand{\lowzlim}{2100}

\newcommand{\lowm}{$4\times 10^{10}\ M_{\sun} < M <10^{11}\ M_{\sun}$}
\newcommand{\him}{$>10^{11}\ M_{\sun}$}

\newcommand{\dumv}{$\Delta (U-V)_z = -0.24 \pm 0.02$ mag}

\newcommand{\dfp}{$\Delta \log_{10} r_e =-0.38 \pm 0.02$}
\newcommand{\mlfp}{$\Delta \log_{10} M/L_B=-0.44 \pm 0.03$}
\newcommand{\dmldz}{$d \log_{10} M/L_B =-0.60 \pm 0.04\ dz$}
\newcommand{\mz}{$z_{\star} = 1.8^{+0.2}_{-0.2}\ $}
\newcommand{\hmz}{$z_{\star} = 2.5^{+0.3}_{-0.2}\ $}

\newcommand{\lowml}{$\Delta \log_{10} M/L_B=-0.48 \pm 0.04$}
\newcommand{\himl}{$\Delta \log_{10} M/L_B=-0.37 \pm 0.02$}

\newcommand{\mlsig}{$\Delta \log_{10} M/L_B=-0.50 \pm 0.03$}
\newcommand{\mlsigt}{ $\Delta \log_{10}  M/L_B = -0.50 \pm 0.02$}
\newcommand{\mlm}{$\Delta \log_{10} M/L_B=-0.47 \pm 0.02$}

\newcommand{\imf}{$x=0.9 \pm 0.2$}
\newcommand{\imfvd}{$x=0.5 \pm 0.2$}

\newcommand{\etal}{{\em et~al.\,}}

\title{$M/L_B$ and Color Evolution for A Deep Sample of $M^{\star}$ Cluster Galaxies  at $z\sim1$:
   The Formation Epoch and the Tilt of the Fundamental Plane
\altaffilmark{1}\altaffilmark{2}\altaffilmark{3}}

\altaffiltext{1}{Based on observations with the NASA/ESA {\em Hubble Space
  Telescope}, obtained at the Space Telescope Science Institute, which
  is operated by the Association of Universities for Research in
  Astronomy, Inc. under NASA contract No. NAS5-26555.  These
  observations are associated with programs nos. 9290, 9772, and 9919} 
\altaffiltext{2}{Some of the data presented herein were obtained at
  the W.M. Keck Observatory, which is operated as a scientific
  partnership among the California Institute of Technology, the
  University of California and the National Aeronautics and Space
  Administration. The Observatory was made possible by the generous
  financial support of the W.M. Keck Foundation.}
\altaffiltext{3}{This paper includes data gathered with the 6.5 m
  Magellan Telescopes located at Las Campanas Observatory, Chile.}

\author{B. P. Holden\altaffilmark{4}}

\author{A. van der Wel\altaffilmark{5}}

\author{D.~D. Kelson\altaffilmark{6}}

\author{M. Franx\altaffilmark{7}}

\author{G. D. Illingworth\altaffilmark{4}}

\altaffiltext{4}{UCO/Lick Observatories, University of California,
  Santa Cruz, 95064, USA; holden@ucolick.org, gdi@ucolick.org}

\altaffiltext{5}{Max-Planck Institute for Astronomy, K\"{o}nigstuhl 17, D-69117, Heidelberg, Germany; 
vdwel@mpia.de }

\altaffiltext{6}{Observatories of the Carnegie Institution of
  Washington, Pasadena, CA, 91101; kelson@obs.carnegiescience.edu}

\altaffiltext{7}{Sterrewacht Leiden, P.O.Box 9513, 2300 RA, Leiden,
  The Netherlands; franx@strw.leidenuniv.nl}

\shorttitle{$L^{\star}$ Cluster Galaxy $M/L$ Evolution}

\begin{abstract} 

  We have measured velocity dispersions ($\sigma$) for a sample of
  \ndisp\ galaxies with $J < 21.2$ or $M_r < -20.6$ mag in \ms , a
  massive cluster of galaxies at $z = 0.83$.  Our data are of
  uniformly high quality down to our selection limit, our 16-hr
  exposures typically yielding errors of only
  $\delta(\sigma)\sim10$\% for $L^*$ and fainter galaxies.  By
  combining our measurements with data from the literature, we have 53
  cluster galaxies with measured dispersions, and {\em
    HST}/ACS-derived sizes, colors and surface brightnesses.  This
  sample is complete for the typical $L^{\star}$ galaxy at $z\sim1$,
  unlike most previous $z\sim1$ cluster samples which are complete
  only for the massive cluster members ($>10^{11}\ M_{\sun}$).  We
  find no evidence for a change in the tilt of the fundamental plane
  (FP).  Nor do we find evidence for evolution in the slope of the
  color-$\sigma$ relation and $M/L_B$-$\sigma$ relations; 
  measuring evolution at a fixed $\sigma$ should minimize the impact
  of size evolution found in other work. The $M/L_B$ at fixed
  $\sigma$ evolves by \mlsig\ between $z=0.83$ and $z=0.02$ or \dmldz,
  and we find \dumv\ at fixed $\sigma$ in the rest-frame, matching
  the expected evolution in $M/L_B$ within 2.25 standard deviations.
  The implied formation redshift from both the color and $M/L_B$
  evolution is $z_{\star} = 2.0 \pm0.2 \pm 0.3 ({\rm sys}) $, during
  the epoch in which the cosmic star-formation activity peaked, with
  the systematic uncertainty showing the dependence of $z_{\star}$ on
  the assumptions we make about the stellar populations.  The lack of
  evolution in either the tilt of the FP or in the $M/L$- and
  color-$\sigma$ relations imply that the formation epoch depends
  weakly on mass, ranging from $z_{\star} = 2.3^{+1.3}_{-0.3}$ at
  $\sigma = 300\ {\rm km\ s^{-1}}$ to $z_{\star} = 1.7^{+0.3}_{-0.2}$ at $\sigma =
  160\ {\rm km\ s^{-1}}$ and implies that the inital mass function similarly varies slowly with
  galaxy mass.

\end{abstract}


\section{Introduction}

The evolution of the mass-to-light ratio ($M/L$) of early-type galaxies
can be measured by combining the velocity dispersion ($\sigma$), the
effective radius ($r_e$), the average surface brightness within the
effective radius ($\langle I_e \rangle$), with which we can measure
$M/L \propto \sigma^2/(r_e \langle I_e \rangle) $ and how it depends on mass, $M
\propto r_e \sigma^2$.  Typically, these three variables are combined
into an empirical relation such as $r_e \propto \sigma^{1.20}
\langle I_e \rangle^{-0.83}$ \citep[][hereafter JFK96]{jfk96}, known as the
fundamental plane \citep[FP][]{faber87,djorgovski1987}.

The absolute $M/L$ value and the how fast the $M/L$ evolves with time
can be used as a technique for measuring the age of stellar population
\citep{tinsley1972}.  The younger the population, the more luminous
the stars, and so it will have lower $M/L$ values and the $M/L$ will
evolve faster.  Massive cluster galaxies, those galaxies with $M >
10^{11}\ M_{\sun}$ where $M \propto \sigma^2 r_e$, out to $z \ge 1$
appear to evolve as $\Delta \ln M/L_B \simeq z $
\citep{vandokkum96,kelson1997,kelson2000c,pvd_mf2001,vandokkum2003,wuyts2004,holden2005,vandokkum2006}. This
rate of evolution is also seen in some field samples, though there is
a larger scatter for the latter
\citep{vandokkum2001b,gebhardt2003,vandokkum2003b,vandeven2003,vanderwel2004,treu2005a,treu2005b,vanderwel2005}.
This rate of evolution implies an epoch of formation for the stars in
early-type galaxies of $z_{\star} \simeq 2$, assuming a passively
evolving simple stellar population with a standard Salpeter-like
initial mass function (IMF). This redshift is the near the peak of star
formation \citep[ e.g;][]{madau98,steidel99,dickinson2003,rudnick2003}.  Thus, the majority
of stars in massive early-type galaxies formed around the time that
the average star in the universe was formed.

The FP results of \citet[][hereafter T05]{treu2005b},
find that for galaxies with masses of $<10^{11}\ M_{\sun}$, the $M/L$
values are lower and evolve more rapidly than those for higher mass
galaxies.  This implies a typical luminosity weighted age of the
stellar populations corresponding to a redshift between $z_{\star}=1$
and $z_{\star}=2$.  \citet[][hereafter vv07]{vandokkum2006} confirmed
that at redshifts of $z\simeq1$, the FP of cluster
galaxies shows a different distribution of $M/L$ values than at
$z\simeq0$ with some lower mass galaxies ($<10^{11}\ M_{\sun}$) having
low $M/L$ and potentially more recent epochs of formation than the
higher mass galaxies.  \citet{jorgensen2006,jorgensen2007} found a
correspondingly steeper tilt to the FP for cluster
galaxies in \cl, indicating that cluster galaxies show similar behavior to field
galaxies, though with a smaller magnitude effect.  Similar results are
clearly seen in \citet{saglia2010} which jointly studies a field and
cluster sample with a uniform selection.  The observed increase in the
tilt of the FP was interpreted as a younger average
stellar population age.  However, for a sample limited in $L$, only
those galaxies with a low $M/L$ will appear at low masses.
\citet[][vdW05]{vanderwel2005} showed that it is difficult to
distinguish between evolution in the tilt and in the scatter of the
FP.  This is the direct result of the selection by optical magnitude
in their sample.  They conclude that the $z\sim1$ FP is likely to be
different from the present-day FP, but that the form (tilt or scatter)
of the evolution is not constrained besides the evolution in the
zero-point at high mass.

Younger stellar populations will not only have a lower $M/L$, but also
have bluer colors. As we look to higher redshift stellar populations
where the slope of the FP appears to tilt, we should see a
corresponding trend of bluer galaxies at lower mass, or a change in
both the slope of the color-magnitude and color-mass relations, with
the caveat that magnitude and mass are not the same
quantity. \citet[][hereafter B06]{blakeslee2005} found that the mean
and scatter in the color-magnitude relation of cluster early-type
galaxies at $z=0.83$ corresponded to a formation epoch of $z_{\star} =
2.2$ over a broad range of magnitudes, and therefore masses.
\citet{mei2009} found no evolution in the slope of the color-magnitude
relation in cluster galaxies out to $z=1.3$ extending the trend found
with lower redshift clusters \citep{stanford98,holden2004}.  This is
complementary to the results of \citet{bell2004} and
\citet{ruhland2009} who found little evidence for evolution of the
slope of the color-magnitude relation of field galaxies.  A lack of
evolution in the slope of the color-magnitude relation when, at the
same time, lower mass galaxies are to apparently have lower mean
$M/L_B$ values is puzzling.

\citet[][hereafter vD08]{vandokkum2008a} combined the $M/L$ evolution
and the evolution in the $U-V$ color of cluster early-types to measure
both the epoch of galaxy formation for massive galaxies and to
constrain the slope of the initial mass function (IMF).  Evolution in
the colors of passively evolving stellar populations are governed by
the location of the main-sequence turn-off, which is determined by the
age of the stellar population.  The rate of $M/L$ evolution, however,
is determined by the slope of the IMF at the turn-off as well as how
bright the stars at the main-sequence turn-off are
\citep{tinsley1972}.  Thus a slow pace of color evolution and a fast
pace of $M/L$ evolution, for example, can be explained by modifying
the initial mass function.  vD08 found, for the more massive cluster
galaxies ($>10^{11}\ M_{\sun}$), a rapid pace of $M/L$ evolution as
compared to the color evolution.  Thus, he concluded that the slope of
the initial mass function at the main-sequence turn-off was flat,
$x=-0.1$ instead of the usual Salpeter $x=1.35$.  This surprising
result did not match previous measurements over the same redshift
range made by other methods \citep{kelson2000c,kelson2001}, a
difference that could be explained by sample selection or the improved
\citet{maraston2005} population synthesis models used by vD08.  This
result is also in conflict with some local indirect measurements of
the IMF through lensing \citep[eg.][]{treu2010,auger2010}, though
possible accommodation can still be made through variation in the dark
matter halos properties.  Because of the IMF, vD08 finds that the
stars formed with a earlier formation epoch, $z=3.7^{+2.3}_{-0.8}$.
vD08 proposes a model that, by $z=2$, predicts a more normal IMF.  The
rapid $M/L$ evolution in the FP of lower mass galaxies implies that
they should have formed at about $z=1-2$.  However, the lack of
evolution in the slope of the color-magnitude relation would imply an
even steeper IMF slope for low mass galaxies than vD08 found for the
high mass galaxies.  This is in marked contrast to the expectations
from the model of vD08 or other IMF slope measurements near $z=2$
\citep{blain1999a}, both of which expect a more normal IMF for galaxies
with masses $<10^{11}\ M_{\sun}$.

Because of the magnitude limits used for selection in previous
work, computing the typical stellar age of formation for $<10^{11}\
M_{\sun}$ galaxies has been challenging.  The highest $M/L$ galaxies,
those with the older populations, cannot be directly observed at
$<10^{11}\ M_{\sun}$ and the impact of these galaxies must instead be
included through a modeling process as was done in T05 or vdW05.  We
selected a sample of $z\simeq1$ early-type galaxies in the field of
\ms, a cluster of galaxies at $z=0.83$ that samples much farther down
the mass function.  The main goal was to search for the more rapid
evolution of the lower mass galaxies implied by the results of other
efforts.  A summary of the sample selection, and our measurements of
the parameters for the FP, the effective radius
($r_e$), the average surface brightness within the effective radius
($\langle I_e \rangle$), and the velocity dispersion ($\sigma$) are discussed in Section
\ref{data}.  We determine the completeness of our sample in Section
\ref{analysis}.  Using these measurements, appropriately weighted, we
measure the FP, color evolution, and the
$M/L_B$ evolution in Section \ref{evol}.  We use these measures to compute
the typical star-formation epoch of the stellar populations and
comment on constraints of the IMF, in Section \ref{results}.  Our results are
summarized in Section \ref{summary}. Throughout this paper, we assume
$\Omega_m = 0.27$, $\Omega_{\Lambda} = 0.73$ and $H_o = 71\ {\rm km\
  s^{-1}\ Mpc^{-1}}$.

\begin{figure*}
\begin{center}
\includegraphics[width=6.2in]{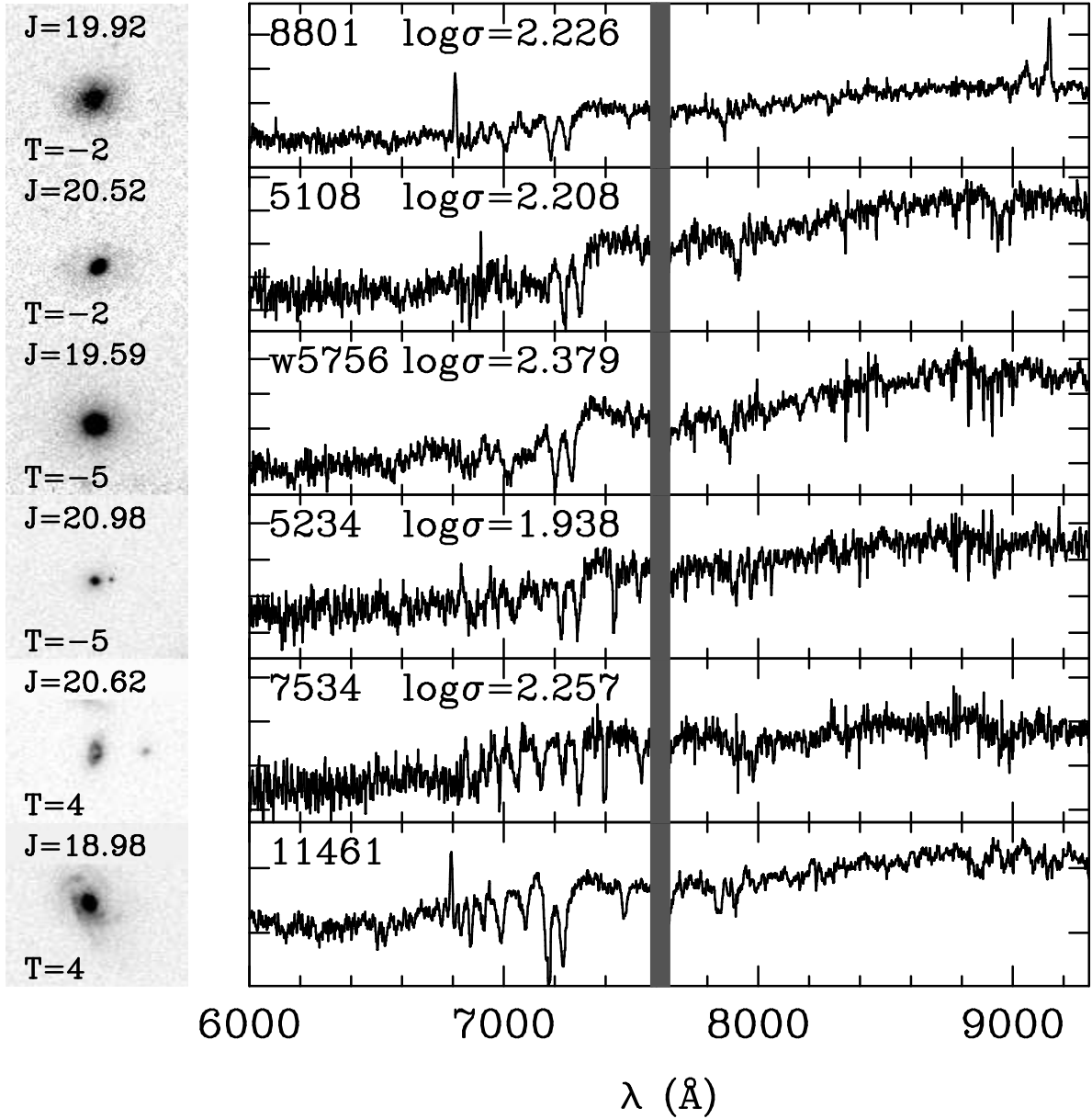}
\end{center}
\caption{ Example spectra and images.  All images are 5\farcs 0 on a
  side and all spectra are in the observed frame.  The vertical gray
  band masks the atmospheric absorption from the $A$ band.  The top
  three galaxies represent typical objects in our sample, $19.5 < J <
  20.5$ with E and S0 morphologies and $\log\sigma > 2.2$.  The brightest
  of these galaxies, w5756, is one of the three galaxies from W04
  which we re-observed.  5234 is the faintest galaxy for which we
  measured a $\sigma$ value.  The bottom two galaxies are both
  late-type systems, with the stretch for 7534 adjusted to illustrate
  this.  7534 has strong enough metal absorption lines for us to
  measure a dispersion.  In 11461 the strong Balmer absorption and
  lack of metal lines prevented us from measuring a $\sigma$, despite
  its bright $J$ magnitude.  Both 5234 and 11461 were imaged using the
  \I\ on ACS while the remaining galaxies were observed with
  \ia\ filter. \label{exam}}
\end{figure*}

\section{Data and Measurements}
\label{data}

\begin{figure*}
\begin{center}
\includegraphics[width=6.4in]{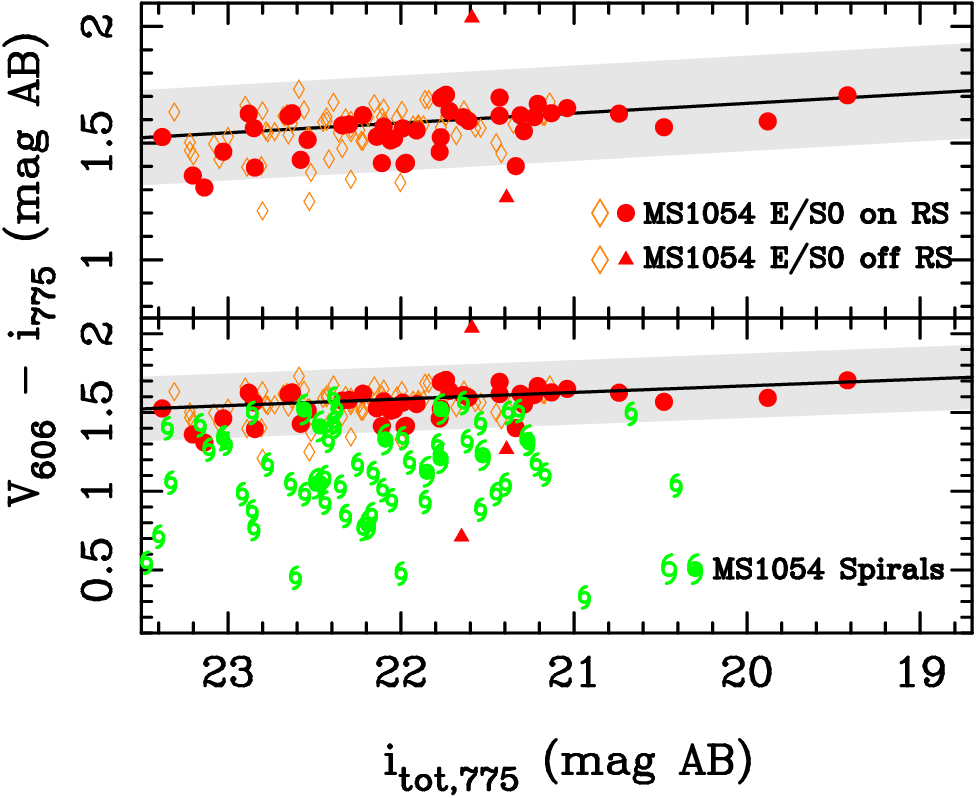}
\end{center}
\caption{ Color-magnitude relation for galaxies in the $z=0.83$
  cluster \ms\ using {\em HST}/ACS data.  We plot the \V$-$\ia\ color
  of galaxies as a function \ia\ magnitude, note different y-axis
  scale for top and bottom panel.  Galaxies for which we have \I\
  data, we convert the \I\ and \V$-$\I\ data into the \ia\ and
  \V$-$\ia\ using the relations in Section \ref{cols}.  Larger red, filled
  circles and triangles are E and S0 galaxies with dispersions while
  smaller orange diamonds are E and S0 galaxies without dispersions.
  Green spirals are spiral or irregular galaxies, with filled spirals
  representing those with measured dispersions.  We show, with solid
  lines, the color-magnitude relation of B06 while the gray regions
  shows the 3$\sigma$ scatter.  We will consider all galaxies within
  that shaded region as on the red-sequence, those galaxies with
  dispersions that lie outside of that region are shown with filled
  triangles.  We find a tight color-magnitude relation for the cluster
  E and S0 population in good agreement with the results from
  B06. This relation extends out to the E and S0 galaxies beyond the
  core of the cluster.  We note that the statistical errors on the
  colors are smaller than the symbols used for plotting, see also
  B06. \label{cm}}
\end{figure*}

We have a sample of \nhst\ galaxies in the field of the $z=0.83$
cluster \ms\ for which we have obtained spectra. Of these, \ndisp\
have dispersion measurements.  For these galaxies we have high
resolution imaging using the Advanced Camera for Surveys (ACS) on the
{\em Hubble Space Telescope (HST)}.  Using the {\em HST}/ACS images,
we measure the sizes and surface brightness by fitting elliptical
S{\'e}rsic models to the data.  The velocity dispersions are measured
by fitting stellar templates broadened by the appropriate Gaussian
line of sight velocity dispersions, representing the typical motion of
stars inside the galaxies.  Below we discuss in detail our
observations and the resulting measurements.

\subsection{$z=0.83$ Sample Definition}
\label{sample}

We based our sample on a magnitude limited survey of galaxies in the
field of \ms.  The original selection was done using a limiting
magnitude of $J=21.2$ mag (Vega) from a catalog of galaxies observed
with the CFH12K at CFHT and the WIRC camera on the du Pont.  Each galaxy in
this sample had either a spectroscopic redshift, or a photometric
redshift from a catalog of $BVRIJH$ imaging data.  Galaxies with
spectroscopic redshifts were preferred in our selection.  These
photometric redshifts, the catalog construction, and the spectroscopic
subsample will be discussed in a later paper.

Beyond meeting the $J$ magnitude limit, the main criterion for
selection was that the galaxy lie in the ACS field of view.  We
selected a DEIMOS mask center and position angle that maximized the
number of galaxies with ACS imaging and with $J<21.2$.  A total of
\nmag\ galaxies met these criteria.  Beyond our main sample, 12
supplemental galaxies were selected to have $21.2<J<21.8$ and an
additional four were targeted outside of the HST field of
view.  Also, we re-observed three galaxies that were in the sample of
\citet[][hereafter W04]{wuyts2004}.  Therefore, the total number
of galaxies targeted for dispersions was \ntot.

We also combine our sample with that from W04.  This earlier sample
has a brighter $I$ selection, see W04 for details, and used LRIS
\citep{oke1995}, with a smaller field of view than DEIMOS, for
spectroscopic observations.  The result is that the combined catalog
of our sample and that of W04 will contain more galaxies in the inner
regions of the cluster and will have a higher completeness at brighter
magnitudes.

Our sample of galaxies is listed in Table \ref{ms1054data} in Appendix
A1.  We list our observed and rest-frame quantities along with the
velocity dispersions.  This table includes the sample of W04, which
are identified by a ``w'' in front of the identification number.

\subsection{{\em HST}/ACS Imaging}
\label{measure}

The ACS imaging used in this paper comes from two separate programs.
The first, discussed in B06, used the F606W, F775W and F850LP filters
to image the center of the cluster. We will refer to the filters as
\V, \ia, and \z\ for the rest of the paper.  The second program imaged
the outer regions of \ms\ with the \V\ and the F814W, or \I\ filter.
Example images can be seen in Appendix A, Figure \ref{exam}.  

We fit a model to each galaxy to determine the effective radius $r_e$
and surface brightness $\langle I_e \rangle$ in the filter that most
closely matches the rest-frame $B$.  For the inner regions, this is
the \ia\ filter while for those galaxies at larger radii, we used \I.
For each galaxy we fit a model with a free S{\'e}rsic parameter
constrained to fall within $1 \le n \le 4$ using {\tt GALFIT}
\citep{penggalfit2002}.  For this paper, we use the best-fitting
S{\'e}rsic parameters for $r_e$ and $\langle I_e \rangle$, as this was
shown to yield the same resulting FP as fixing $n=4$
\citep{kelson2000a}.  When the best fitting S{\'e}rsic $n$ was at the
limit of the allowed range, i.e.\, $n=1$ or $n=4$, we refit the image
fixing $n$ to that value.

The total magnitude we use is the normalization of the model fit.  To
measure colors, we used the circularized half-light or effective
radius, $r_e = a_{hlr} \sqrt q$, where $q$ is the ratio of the minor
to major axis, or $1-\epsilon$, and $a_{hlr}$ is the half-light radius
along the major axis of the best fitting elliptical model as
determined by {\tt GALFIT}.  This is the radius used both in our $M/L$
analysis and the radius of the aperture for which we measured
the galaxy's color, see B06 for details.  From this circularized $r_e$, we
compute the average surface brightness within that radius, or $\langle I_e \rangle $.

Using the simulation results of \citet{holden2009}, which placed real
galaxy images in the ACS frames, we find that our measurements of the
total magnitudes are too bright by $-0.08$ mag in the \ia\ data or
$-0.04$ mag in the \I\ data.  We find a scatter of $\sigma = 0.07$
mags around the input magnitude for the simulations, after removing
the offset, in good agreement with the expectations from total
magnitude measurements in ACS imaging \citep{holden2005b}.  We apply
this offset to our data in later sections.

In addition to fitting a bounded S{\'e}rsic model, we fit a fixed de
Vaucouleur's model to our data so we can compare the two resulting
fits.  As expected, for galaxies classified as E or S0 the median best
fitting $n$ value was $n=3.6$ while, for those galaxies classified as
spirals the median $n$ was $n=1.04$ (we discuss the classifications in
the next section.)  The $n$ values showed a larger scatter for the
spirals than for E and S0 systems as well.  In cases where the best
fitting $n>3$, the values for $r_e$ are good agreement with the $n=4$
values, with differences of 11$\pm$6\% on average.  When best fitting
bounded S{\'e}rsic $n<3$, the differences in the $r_e$ are much
larger, as expected \citep{kelson2000a}.

\subsubsection{Morphologies}
\label{morph}

Galaxy morphologies were obtained from the literature for all the
galaxies in our sample.  All of the morphologies were determined by
Marc Postman from a $J$ selected catalog with a limit of $J=22.5$ mag.
Each galaxy was classified in the passband closest to the rest-frame $B$,
either the \ia\ or the \I.  The galaxies were classified in the same
manner as those in \citet{postman2005}, and the galaxies in the
central regions of \ms\ were classified twice, once for this paper and
once for the original \citet{postman2005} study.  The scatter for the
fraction of E+S0 galaxies was 6\%, the same as was found by using
multiple classifiers in \citet{postman2005}.  The morphologies, using
the T system of \citet{postman2005}, are listed in Table
\ref{ms1054data}.  Galaxies with $T<0$ are E and S0 systems.

\subsubsection{Redshifted Magnitudes}
\label{kcorr}

We redshift the $U$, $B$ and $V$ passbands to z=0.83 and use the
observed photometry to compute what magnitudes we would observe in
these redshifted passbands.  Hence, these passbands measure the flux
densities at the redshifted effective $\lambda$ for each filter;
$\lambda_{U}(1+z)$, $\lambda_{B}(1+z)$ and $\lambda_{V}(1+z)$.  For
brevity, we will call these redshifted magnitudes and label these
magnitudes $U_z$, $B_z$ and $V_z$.  With the removal of the additional
dimming caused by the distance to the galaxies, these magnitudes would
become rest-frame magnitudes.  We compute these transformations
between observed and redshifted magnitudes using a similar process as
B06, \citet{holden2006} and \citet{holden2007}, but based on the same
basic approach as \citet{vandokkum96} and \citet{kelson2000a}.

Both the \ia\ and \I\ filters are mapped into $B_z$.  We computed
magnitudes at both $z=0$ and $z=0.83$ for a set of \citet[BC03]{bc03}
$\tau$ model templates.  For the $z=0$ filters, we used the $U$, $B$
and $V$ curves from \citet{buser1978}, specifically the B3 curve for
the $B_z$ as tabulated by BC03.  We note that the transformations for
the ACS passbands \V, \ia\ and \z\ all match closely the
transformations from B06.  We explicitly note here that we do not
attempt to use the BC03 templates to model the photometry directly.
Rather, we use the templates to compute a grid of observed and
redshifted magnitudes to which we fit a simple linear or quadratic
relation.

The $\tau$ models used for the transformations span a range of
parameters.  These models had exponential time-scales of 0.1 to 5
Gyr, ages from 0.5 Gyr to 12 Gyr and three metal abundances, 2.5
solar, solar and 0.4 solar.  We then fit a linear or quadratic
relation between the observed color or magnitude and the redshifted
magnitude or color, for all models including all ages of those models.
For the ACS transformations, we restricted the transformations to the
range in colors we observed for galaxies at the redshift of the
cluster and tabulate them in Table \ref{transform}.  All observed
magnitudes use the AB system unless otherwise noted.

\begin{deluxetable*}{lrl}
\tablecolumns{3}
\tablecaption{Transformations from Observed to Redshifted Magnitudes \label{transform}}
\tablehead{
\colhead{Filter\tablenotemark{a}} & \colhead{Transformation}  &
  \colhead{Color Range} \\ 
\colhead{(mag)} & \colhead{(mag AB)} & \colhead{(mag AB)} \\
}
\startdata
$r_z$ &  $= J + 0.118 (I-J) +  1.538$\tablenotemark{b} &  \\ \hline
 $B_z$ &  $= i_{775} - 0.113 (V_{606}-z_{850}) + 0.827$ &
 $(V_{606}-z_{850}) < 1.5$ \\
 $B_z$ & $= i_{775} - 0.190 (V_{606}-z_{850}) + 0.964$ & $1.5 <
 (V_{606}-z_{850}) < 2.6$ \\
 $(U-V)_z$ &  $= 0.876 (V_{606}-z_{850}) - 0.757$ &  $(V_{606}-z_{850}) < 1.5$ \\
 $(U-V)_z$ & $= 1.002 (V_{606}-z_{850}) - 1.063$  & $1.5 <
 (V_{606}-z_{850}) < 2.6$ \\
$B_z$ & $= I_{814} - 0.003 (V_{606}-I_{814}) + 0.778$ & $V_{606}-
I_{814} < 1.2$ \\
$B_z$ & $= I_{814} - 0.006 (V_{606}-I_{814}) + 0.783$ &$1.2 < V_{606}
- I_{814} < 2.4$ \\
$(U-V)_z$ &  $=  1.135 (V_{606}-I_{814}) - 0.861$ & $V_{606}- I_{814} <
1.2$ \\
$(U-V)_z$  & $=  1.428 (V_{606}-I_{814}) - 1.363$ & $1.2 < V_{606}
- I_{814} < 2.4$ \\
\enddata

\tablenotetext{a}{Results are in Vega magnitudes for the $UBV$ transformations,
  and we used the \citet{buser1978} filter curves.  For the $r_z$, the
 magnitudes are AB and use the SDSS filter curves.}
\tablenotetext{b}{The observed magnitudes are in the Vega system,
  unlike the other photometry in this table.}

\end{deluxetable*}

\subsection{Spectral Observations}
\label{specobs}

The spectral data were acquired over two observing runs in 2008 January and
March, in excellent seeing and generally photometric
conditions.  The exposures were 20 minutes long using the 600 line
mm$^{-1}$ grating on the DEIMOS spectrograph.  The slits were
1\arcsec\ wide, yielding a resolution of $\sim$3.8 \AA\ or an
instrumental $\sigma=$60 km
s$^{-1}$.  The slits were at least 8\arcsec\ long, and between each exposure we
offset the telescope by 2\arcsec\ along the slit.  Therefore, the
object does not land on the same detector pixels in sequential
exposures, but is instead offset by 2\arcsec.  Each mask was observed
for six exposures, or w hr.  The final exposure was 16 hr for
the deepest galaxies, 10 hr for the brighter objects.

\subsubsection{Spectral Data Reductions}
\label{specred}

Our sky subtraction and spectral extraction used the techniques
outlined in \citet{kelson2003} as implemented by D.\ D.\ Kelson.  We
used the internal quartz lamps to flat field the data.  Wavelength
solutions for each observation were determined using the night sky
lines.  Each slitlet has an independent sky model, see
\citet{kelson2003} for details, which is then subtracted
from each 20 minute observation.

Bright stars in each mask are used to measure the spatial offset
between frames, as we offset the telescope between each exposure.
After the sky model was subtracted from each frame, we then subtracted
sequential frames to remove any residual sky signal.  Each spectrum in
each frame has a separate model used for optimal extraction
\citep{horne1986}.  The initial guesses for the profiles of the
spectra are based on the bright stars in each mask, but are then fit
to each spectrum in each exposure.  The final one dimensional spectrum
is an extraction based on  all of the separate exposures simultaneously,
re-binned to the typical dispersion of the data, 0.64 \AA\ pixel$^{-1}$.
We plot example spectra in Appendix A, Figure \ref{exam}.

\subsubsection{Measuring the Velocity Dispersions}
\label{veldisp}

Velocity dispersions and high-precision redshifts are measured with a
direct fitting method \citep[see, e.g.,][]{kelson2000b}, as described
by \citet{vanderwel2004} and vdW05, using a high-resolution solar
spectrum that is smoothed and re-binned to match the galaxy
spectra. If the continuum of the template and galaxy spectra are well
matched, template mismatches from different spectral types introduce
errors of, 2\% -5\%. This, combined with a small systematic uncertainty
in the wavelength calibration and the possible non-Gaussian velocity
distribution of the stars, prompts us to add in quadrature an
empirically determined 5\% error (see vdW05 for a fuller explanation)
to the formal fitting error on the measured velocity
dispersion. Fitting in real space, as opposed to Fourier space, has
the advantage that pixels can be weighted by the signal-to-noise (S/N)
ratio. In addition, we always mask regions that are affected by the
atmospheric A and B absorption bands and emission lines. By default,
we also mask Balmer absorption features, as these are unsuitable for
velocity dispersion measurements. However, the H$\epsilon$ line, which
is blended with the Ca H line, is not masked if Balmer absorption
lines are weak, as is the case for the majority of the sample for
which we managed to measure velocity dispersions (26 out of 39,
including the three from Wuyts \etal\ 2004 that were re-observed).  For
the remaining 13 galaxies, Balmer lines are the dominant absorption
features, such that we are forced to mask the Ca H and H$\epsilon$
blended line as well. The final velocity dispersions include a 6.6\%
aperture correction to give the central velocity dispersion within a
fixed physical aperture of 1.6 kpc or 3.4\arcsec\ at the redshift of
Coma, the same aperture used by JFK96.

We note that thanks to the excellent quality of the spectra, only two
galaxies out of the 55 targeted cluster galaxies with ACS imaging do
not have measured velocity dispersions because of too low S/N in the
continuum. For a further 14 galaxies we cannot determine the velocity
dispersion due to the lack of metal features, as their continuua are
either featureless (with emission lines) or dominated by strong Balmer
absorption lines. These are all classified as late-type galaxies, in
agreement with their spectral characteristics. Finally, one early-type
galaxy has no measured velocity dispersion because of a moderate
active galactic nucleus contribution. Because of this high success
rate, we now have, for the first time, a high-redshift sample of
early-type galaxies which spans a factor of $\sim$4 in velocity
dispersion and 2 orders of magnitude in dynamical mass.  All
galaxies with spectra are listed in Table \ref{ms1054data}, those
without measured dispersions are so marked.

\subsection{Low-redshift Comparison Sample}
\label{lowz}

Our low redshift comparison sample is a subset of the Coma sample from
\citet{holden2007}.  Each galaxy has surface brightness profiles fit
to the Sloan Digital Sky Survey (SDSS) imaging data in the same manner
as we have for \ms .  Each galaxy is a member of the JFK96 E and S0
sample used in that FP study.  We use the dispersions from JFK96 and
the colors from \citet{eisenhardt2007}.  The colors from Coma match
the \citet{ble92b} colors well, but cover a larger number of galaxies.
We note here that we use only the half-light radius colors from Table
8 of \citet{eisenhardt2007}.  For the total magnitudes, we use our
values from the surface brightness profile fits to the SDSS images.
The larger sample of sizes and colors allows us to expand on the sample
of JFK96 which had $B$ data for only a subset of all of the Coma
galaxies with dispersions.  We list in Table \ref{comadata} the
magnitudes, rest-frame colors and dispersions we use.  Not all of the
sample of JFK96 have colors listed in \citet{eisenhardt2007} as noted
in Table \ref{comadata}.

We use the same procedure and $\tau$ model population templates from
BC03 as we used for the \ms, see Section \ref{kcorr}, to transform the
$z=0.023$ observed $U$ and $V$ into $z=0$ values $U_z$ and $V_z$.  We
find that there is some curvature to the relation, and we require a
quadratic fit of
\[ (U-V)_z = (U-V) -0.067 (U-V)^2+ 0.220 (U-V) -0.220\] to transform
the observed magnitudes into the redshifted ones.  For galaxies on the
red-sequence of Coma, we find $(U-V)_z = (U-V)-0.04$ mag.  Using the
\citet{cww80} elliptical model, we find $(U-V)_z = (U-V) -0.04$ mag,
in good agreement with our quadratic relation.  We note that these
values are significantly different from the $(U-V)_z = (U-V) - 0.00$
as found by \citet{ble92b}.  For the rest of the analysis, we will use
the second order equation above relating the observed $U-V$ and the
$z=0$ $(U - V)_z$ color.

\section{Data Analysis}
\label{analysis}

\subsection{The Colors of the Galaxy Population}
\label{cols}

The colors of the galaxy population in \ms\ have been studied extensively
in B06.  We show in Figure \ref{cm} the color-magnitude diagram of the
population of galaxies that we targeted for dispersions in this study
and the larger sample of galaxies that could have been targeted, a
superset of the B06 sample.  We note here that we remeasured the
colors for all galaxies in our sample, instead of using the values
from B06 and we will use these colors for our later measurements of
the evolution of the cluster population.

Because two different programs observed \ms\ which used two different
sets of filters, we convert the colors and magnitudes of galaxies
observed in \I\ to \ia\ using the following relations
\[ i_{775} = I_{841} + 0.248 ( V_{606} - I_{814}) - 0.211 \],
\[ V_{606} - i_{775} = 0.752 ( V_{606} - I_{814}) + 0.211 \], which are
only valid between $1.2 < V_{606} - I_{814} < 2.4$.  In Figure
\ref{cm}, we also removed galaxies from our sample that do not have
spectroscopic redshifts.

Our colors significant a strong color-magnitude relation with a small
scatter.  The scatter for the whole sample of E and S0 galaxies with
redshifts is $\sigma = 0.077 \pm 0.008$, larger than the scatter
measured in B06.  This larger scatter comes from the subsample of
galaxies with \V$-$\I\ colors that have been transformed into
\V$-$\ia\ colors, removing them brings the scatter down to the size as
measured in B06.  This is consistent with the $\sim$0.01-0.02 mag
error typically found in these color transformations.  We find we recover the
slope and intercept from B06 with our larger sample, showing
consistency between the two sets of measurements.

\subsection{Success Rate}
\label{success}

Evaluating the degree of evolution in the galaxy population requires
calculating the completeness of our sample.  Part of this is
estimating how successful we are measuring dispersions as a function
of galaxy property.  We targeted a total of \ntot\ galaxies, with
\nmag\ above of magnitude limit of $J<21.2$.  \ndisp\ have measured
dispersions, with \ndispm\ above $J<21.2$.

The simplest way of examining the completeness is to plot the
distribution of galaxies with measured dispersions as a function size,
color, and magnitude.  First we show the color-magnitude plot in Figure
\ref{colmag}, followed by the size-magnitude relation in Figure
\ref{sizemag}.  In both cases, the magnitude is the $J$, the selection
passband.

\subsubsection{The Success Rate of Measuring Dispersions for E and S0 Galaxies}

\begin{figure*}
\begin{center}
\includegraphics[width=6.4in]{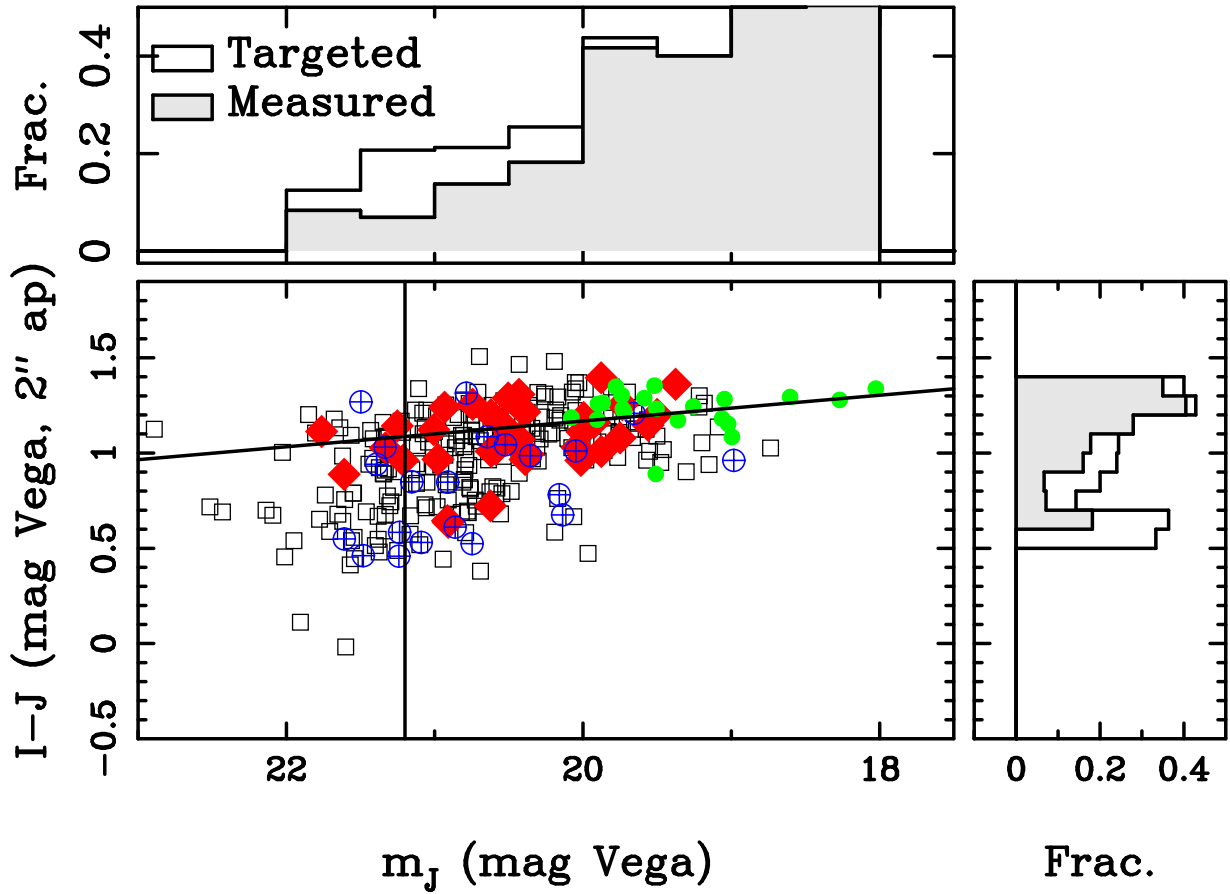}
\end{center}
\caption{Color-magnitude relation for the parent sample of galaxies in
  \ms.  We plot the $I-J$ color of galaxies as a function $J$
  magnitude.  The solid diamonds are the galaxies with measured
  dispersions from this paper while open circles with ``$+$'' symbols
  are galaxies we targeted but did not measure a dispersion for.
  Solid circles are galaxies with dispersions from W04.  We show, as a
  vertical line, our magnitude limit of $J=21.2$.  The other line is
  the color-magnitude relation fit to the E/S0 population. Along the
  top, we show histograms of the fraction of galaxies with dispersions
  (filled gray) and targeted (open) as a function of $J$ magnitude.
  This includes the galaxies from W04 - or the solid circles - so this
  plot shows the relative weight of the different galaxies in our
  analysis.  Along the right side, we plot the fraction of galaxies
  with dispersions and targeted with $J<21.2$ by color.  We find that
  our sample of galaxies with dispersions covers most of the range of
  colors and magnitudes in the parent sample.  We successfully measure
  dispersions for $\sim$80\% of E/S0 galaxies and $\sim$50\% of spiral
  galaxies above our $J=21.2$ magnitude limit that we targeted.  Thus,
  26\% of all cluster galaxies in the ACS imaging and with $J<21.2$
  have dispersions, matching the visual impression of this figure.
  For 14 galaxies we targeted, we cannot determine the velocity
  dispersion due to the lack of metal features, as their spectra are
  either featureless (with emission lines) or dominated by strong
  Balmer absorption lines, producing a decreasing fraction of
  galaxies with bluer $I-J$ colors. \label{colmag} }
\end{figure*}

\begin{figure*}
\begin{center}
\includegraphics[width=6.4in]{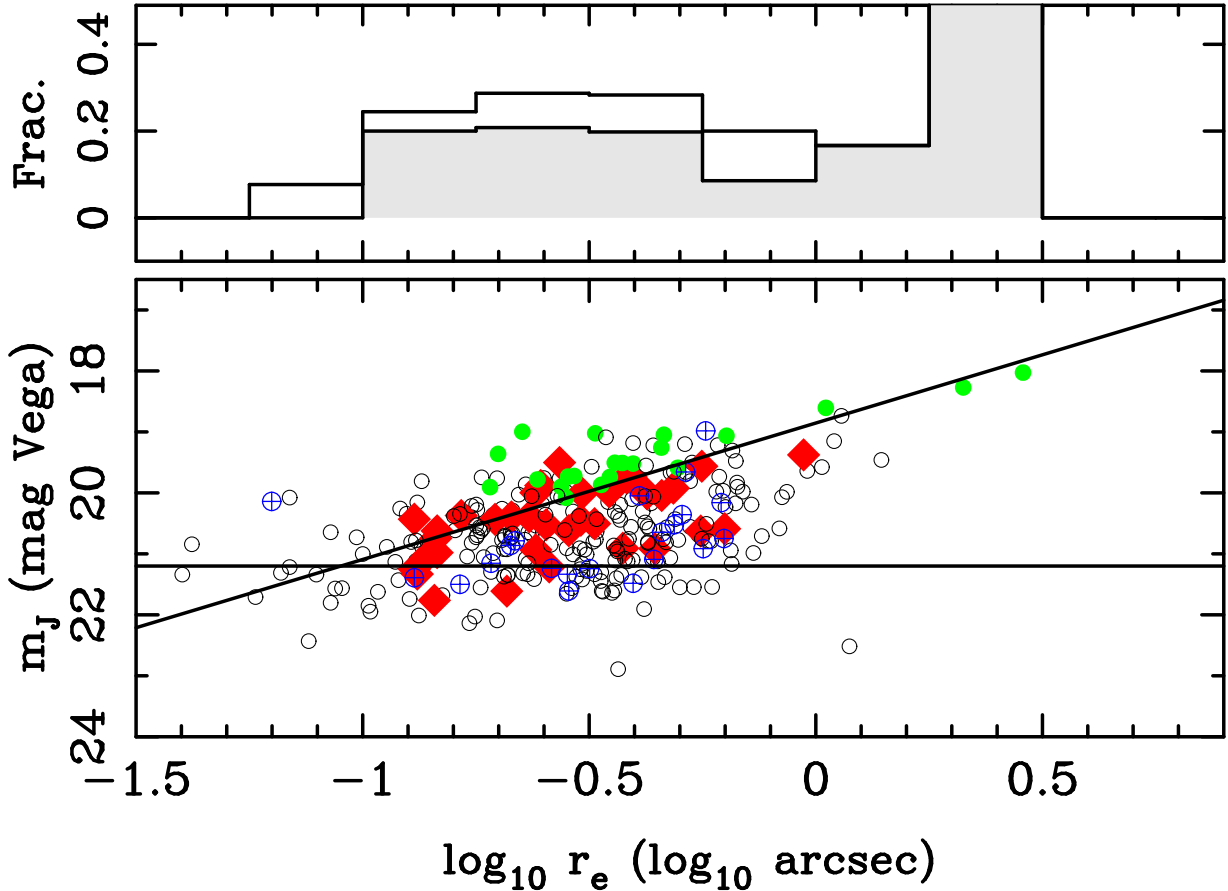}
\end{center}
\caption{Size-magnitude relation for the parent sample of galaxies in
  \ms.  We plot the $J$ magnitude of galaxies as a function {\em
    HST}/ACS measured half-light radii using the same symbols as
  Figure \ref{colmag}.  The fraction of galaxies with dispersions and
  $J<21.2$ is plotted as a function of size in the gray histogram
  above the plot, and the fraction of galaxies targeted for dispersion
  measurements is shown with the open histogram.  We find that
  fraction of galaxies with dispersions is relatively flat with size
  above our magnitude limit, with exception of the two brightest
  galaxies in the cluster and the smallest galaxies which are
  predominately below are magnitude limit.  This means that we have no
  strong surface brightness selection bias.  This is in accordance
  with previous work which finds that the S/N of a spectrum is
  proportional to $\langle I_e \rangle r_e^{1.9}$, which is close to $\langle I_e \rangle r_e^2$ or
  the observed magnitude
  \citep[T05;][]{vanderwel2008c}. \label{sizemag} }
\end{figure*}

We compute the fraction of galaxies with measured dispersions as a
function of the magnitude and offset from the color-magnitude relation,
shown in Figure \ref{colmag}.  Our success in measuring velocity
dispersions for E and S0 galaxies was high, with 29 dispersions out of
the 34 galaxies targeted.  The rate of success for measuring
dispersions ranges from 100\% at the brightest magnitudes to 70\% for
the galaxies below our magnitude limit $J=21.2$, averaging $\sim$80\%
above our magnitude limit.  When we look at galaxies as a function of
$I$ instead of $J$ magnitude, we find a similar distribution.  The
fraction of galaxies targeted with measured dispersions is flat for
$I<22$ at 93\%, dropping to 70\% for the faintest $22.5<I<23$.

For E and S0 galaxies, we find no evidence of a size dependence on the success rate
of measuring dispersions.  This likely implies that magnitude is the
important variable in deciding the success of a dispersion
measurement.  Both T05 and \citet{vanderwel2008c} found that the signal
to noise of a spectrum was predicted by the scaling of FP
variables $\langle I_e \rangle r_e^{1.9}$, which is very close to magnitude of a
galaxy, $\langle I_e \rangle r_e^{2}$.

\subsubsection{The Success Rate of Dispersions for Early-type Spiral Galaxies}

Our success with late-type galaxies is lower, with 10 measured
dispersions out of the 20 targeted galaxies with $J<21.2$, or 50\% on
average.  We find that the success rate is flat with size and
S{\'e}rsic index.  It is also flat with magnitude, but only for
galaxies $J<21.2$.  We measure no dispersions for the spiral galaxies
below our $J<21.2$ magnitude limit.  The most important variable is
the color, as we are biased toward measuring the dispersions of red
galaxies.  Early-type spirals on the red-sequence have a 100\% success
rate (see Figure \ref{cm} for our definition of the red-sequence),
while blue-ward of that, the rate drops to $\sim$30\%.  Because of
this strong color dependence, we will give bluer spirals galaxies
higher weight in our later analysis.  Measuring the velocity
dispersion requires the presence of strong metal lines, so we are
biased against young or low metallicity clusters in our sample.  The
small spiral fraction in \ms\ for moderate mass galaxies
\citep{holden2007} means the overall impact of this bias will be
small.

\subsection{How Representative is the Sample?}
\label{wgt}

Thirty percent of the galaxies with $J<21.2$ that could have been targeted had
slits placed on them and 26\% of galaxies with $J<21.2$ have measured
dispersions. These numbers, of course, depend mildly on the apparent
magnitude of the galaxy.

The second sample we would like to consider is the published work of
W04.  W04 had a selection limit of $I=22$ but preferentially targeted
brighter galaxies, with more than half of the sample above $I\le 21$
(or $J \le 19.8$).  This results in a sample with a much brighter
effective magnitude limit.  As can be seen in Figure \ref{colmag},
most of the galaxies that could have been targeted, and the ones that
had been targeted, are on the red-sequence.

Considering only E and S0 galaxies, our spectroscopic sample, combined
with that of W04, targeted 38\% of the galaxy population with
$J<21.2$, with 36\% of E and S0 galaxies above $J=21.2$ having
dispersions.  This combined sample of velocity dispersions represents
a significant fraction of the E and S0 population in \ms.  For the
rest of the paper, we will use the completeness fraction of galaxies
as a function of $J$ magnitude as weights.  The inclusion of the W04
sample makes these weights a strong function of magnitude, as can be
seen in Figure \ref{colmag}.  These completeness fractions range from
100\% at the brightest magnitudes to 20\% at our completeness limit of
$J=21.2$.

\section{Evolution of the Fundamental Plane and Galaxy Properties}
\label{evol}

\subsection{Fundamental Plane Evolution}
\label{fpevol}

\subsubsection{Measuring the Tilt of the Fundamental Plane}

We compute the parameters of the FP by fitting the
$J<21.2$ magnitude limited subset of our dispersions ($\sigma$),
effective radii ($r_e$) and our $B_z$ surface brightnesses.  To fit
the FP, \( \log r_e = \alpha \log \sigma + \beta \log
\langle I_e \rangle + \gamma\), where $\langle I_e \rangle$ is $-0.4 \mu_B$, we minimized the mean absolute
orthogonal deviation, as was done in JFK96:
\[ \Delta = \frac{\log r_e - \alpha \log \sigma - \beta \log \langle I_e \rangle -
  \gamma}{\sqrt(1+\alpha^2 + \beta^2)} \] around the FP
in all three projections, $r_e$, $\sigma$ and $\langle I_e \rangle$.  We computed the
average values of each of the coefficients for our final estimates of
$\alpha$, $\beta$ and $\gamma$.  We estimated the errors on all of our
fits to the FP by bootstrapping the data.

To compute the best fitting plane, we need include our incompleteness.
We did this two different ways.  First, we use our weights we computed
above in Section \ref{wgt}.  Thus, galaxies at the fainter end of the
luminosity function, closer to our limiting magnitude, get a higher
weight.  Second, we trim the data at a fixed value of $\sigma$.
Because galaxies become intrinsically fainter at lower masses, by
trimming at a fixed $\sigma$, we remove the very lowest $M/L$ galaxies
where our sample will not have the corresponding high $M/L$ galaxies
at the same dispersion.  We find that at $J=21.2$, the typical
dispersion is $\log_{10} \sigma = 2.10 \pm 0.10$, and thus we select
limit of $\log_{10} \sigma = 2.20$.  We will examine how our best
fitting plane varies depending on whether or not we remove the lowest
$\sigma$ galaxies or use our weights.

Our resulting best fit values are summarized in Table \ref{FPvals}
and, for comparison, we give the relations from JFK96 for the $B$
selected sample.  We verified that we recover the FP values from JFK96
when we fit to the values listed in Table \ref{comadata}.  

When fitting the FP in \ms, at first we restricted the sample to only
those galaxies that are E and S0 galaxies, a similar selection as
JFK96, but using the $\log_{10} \sigma > 2.2$ and using the weights
from above, we find $\alpha=1.25 \pm 0.11$ and $\beta=-0.76 \pm 0.03$.
The value of $\alpha$ is in good agreement with that of JFK96 but the
value of $\beta$ appears different at the level of two standard
deviations.  \citet{hyde2009b} show that sample selection can cause
large changes in the best fitting slopes of the FP, and
specifically trimming at a fixed $\sigma$.  The JFK96 sample is
complete to a fainter $r$ limit than the equivalent $J$ limit for
our sample and did not have a limiting $\sigma$ as we did.  So, we
restricted the sample of Coma galaxies to those galaxies that would
match our high redshift selection criteria, namely E and S0 galaxies
with an equivalent $r$ selection corresponding to our $J<21.2$ at
$z=0.83$ and the same limiting $\sigma$.  We refit the FP for the Coma
sample, and we find that value of $\beta$ for the Coma galaxies
decreases to $\beta = -0.78 \pm 0.04$ with $\alpha$ unchanged.  Thus,
we conclude that the mild difference in $\beta$ we find in comparison
to JFK96 is likely a result of sample selection, and that there is no
evolution in the tilt of the FP.

Including early-type spirals in our sample does not change the result
significantly, as was seen in a $z=0.33$ cluster by
\citet{kelson2000c}.  Fitting the FP to all galaxies
with $\log_{10} \sigma > 2.2$ and using the weights from above, we
find $\alpha=1.18 \pm 0.16$ and $\beta=-0.76 \pm 0.06$.  Lowering the
limiting $\sigma$ to $\log_{10} \sigma > 2.1$ and using the weights
from above, we find $\alpha=1.19 \pm 0.13$ and $\beta=-0.78 \pm 0.05$.
In general, these numbers are in good agreement with the values for a
more restrictive subsample of E and S0 galaxies with $\log_{10} \sigma
> 2.2$.  This shows that our results are robust to different limiting
velocity dispersions and uncertainties in the sample selections.

\begin{deluxetable*}{lccll}
\tablecaption{Summary of FP Parameters \label{FPvals}}

\tablecolumns{5}
\tablehead{
\colhead{Cluster} & 
\colhead{Limiting $\log \sigma$} & \colhead{Morphology} &
\colhead{$\alpha$} & \colhead{$\beta$} \\
\colhead{} & \colhead{(${\rm km\ s^{-1}} $)} & \colhead{} &\colhead{} & \colhead{}  \\
}
\startdata
JFK96\tablenotemark{a} &  & E and S0 & $1.20 \pm 0.07$ & $-0.83
\pm 0.02$  \\ 
Coma\tablenotemark{b}   & 2.2& E and S0 & $1.18 \pm 0.08$ & $-0.78 \pm 0.04$  \\ \hline
\ms & 2.2 & All &  $1.18 \pm 0.16$ & $-0.76 \pm 0.06$  \\
\ms & 2.2 &  E and S0 & $1.25 \pm 0.11$ & $-0.76 \pm 0.03$ \\
\ms & 2.1 & All &  $1.19 \pm 0.13$ & $-0.78 \pm 0.05$  \\

\enddata
\tablenotetext{a}{The best fitting slopes for the $B$ from the $r$ selected Coma
  sample from Table 5 of JFK96.}
\tablenotetext{b}{The best fitting slopes as determined by for the
  Coma sample from Table \ref{comadata} after selecting galaxies above
  the $\log \sigma$ limit and above the equivalent $r$ magnitude limit
  that matches our $J$ magnitude limit at $z=0.83$.}

\end{deluxetable*}

\begin{figure*}
\begin{center}
\includegraphics[width=6.4in]{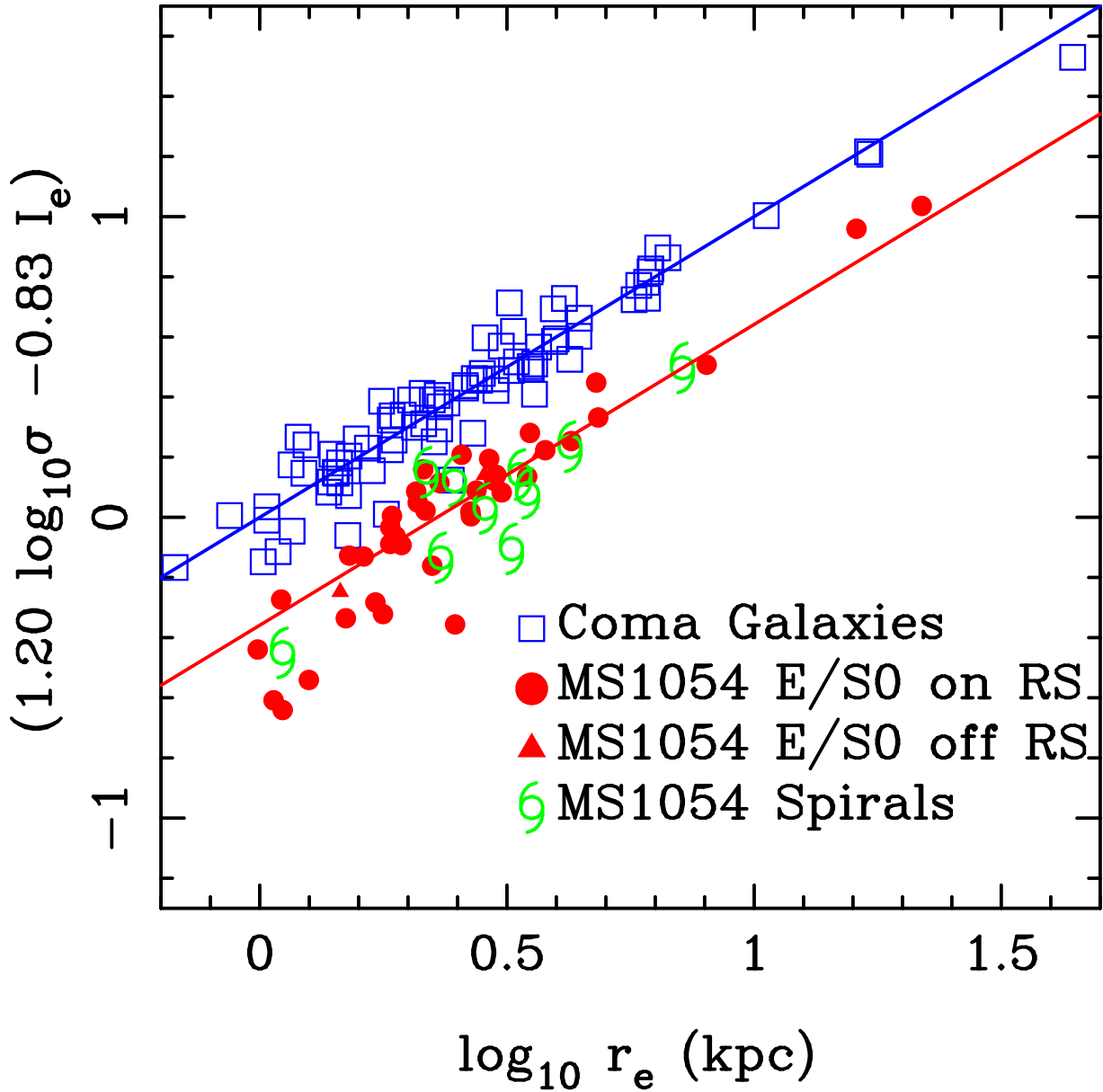}
\caption{Fundamental plane for the Coma sample of JFK96 (blue open
  squares) and our sample in \ms.  Spirals are marked by green spiral
  symbols while E and S0 galaxies are filled red points, circles for
  those on the red-sequence and triangles for those off.  We plot all the
  galaxies in Coma, regardless of $\sigma$ or $r$ magnitude.  We use
  the values of the slope of the FP, $\alpha=1.20$ and
  $\beta=0.83$, from JFK96, and plot the FP relation as a blue solid
  line for Coma.  The red solid line is the FP relation for \ms\ using
  the same slope as we do for Coma, but with an offset to match the
  relation in \ms.  We find no evidence that the tilt in the
  FP evolves, as we recover the relation of JFK96 for
  the galaxies in our sample.  We do find that the FP has a different
  zero-point.  The change in the zero-point is \dfp, which corresponds
  to evolution of \mlfp\ for the stellar populations of the cluster
  galaxies.  \label{fp}}
\end{center}
\end{figure*}

From these tests we conclude that the observations are consistent with an
unchanging tilt in the FP.  In Figure \ref{fp}, we plot the
FP for our $z=0.83$ sample.  We use the values of
$\alpha$ and $\beta$ from JFK96 to plot both the low and high redshift
samples.

\subsubsection{Residuals Around the Fundamental Plane}
\label{resid}

vv07 found that there was a strong correlation between the residuals
around $\log r_e$ and other quantities related to the mass of the
galaxy, such as $\sigma$ or the dynamical mass.  Such a deviation
could be caused by curvature in the FP and is more
thoroughly explored in \citet{vandermarel2007a,vandermarel2007b}.  We
assume the FP relation from JFK96 and we plot, in
Figure \ref{devs}, the deviation between the measured $r_e$ and the
predicted $r_e$ from the FP as a function of $r_e$,
$\sigma$, and the dynamical mass $M_{dyn}$.  We computed $M_{dyn}$ is
in $M_{\sun}$ using the relation \( M_{dyn} = 5\sigma^2 r_e / G\)
or \[ \log_{10} M = 2 \log_{10} \sigma + \log_{10} r_e + 6.07 \] with
$\sigma$ is in km s$^{-1}$ and $r_e$ is kpc..  We find galaxies with
dynamical masses $M_{dyn} <10^{11}\ M_{\sun}$, that lie close to the
$z=0$ relation after adjusting for passive $M/L$ evolution (see
below).  This result is expected given the lack of evolution we find
in the tilt of the FP.

\subsubsection{$M/L$ Evolution from the Offset of the Fundamental
  Plane}

\begin{figure*}
\begin{center}
\includegraphics[width=6.0in]{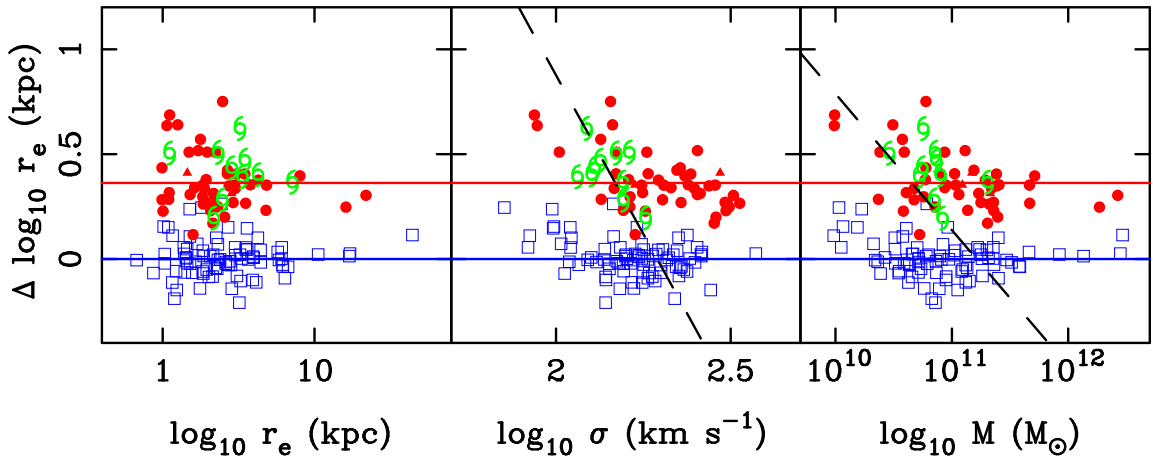}
\end{center}
\caption{Deviations from the FP of JFK96 as a function
  of $r_e$, $\sigma$, and mass ($M_{dyn} = 5 \sigma^2 r_e / G$).  We
  use the same symbols as in Figure \ref{fp} and show the approximate
  $J=21.2$ selection limits for a red-sequence galaxy at $z=0.83$ with
  dashed lines.  Galaxies below those lines are in our
  fainter subsample ($21.2 < J < 21.8$).  The average offset of the
  FP, \dfp\ (which corresponds to \mlfp), is
  illustrated with a red line.  The offset of $z=0.83$ galaxies from
  the $z=0.023$ FP is not larger for galaxies with low
  $\sigma$ or mass as long as they lie above our limiting magnitude,
  as expected if the slope or tilt of the FP does not
  evolve. \label{devs} }
\end{figure*}

Traditionally, the offset in the FP is used to measure
the amount of $M/L$ evolution.  This is done by assuming that the
offset, the value of $\Delta \log_{10} r_e$, is all caused by
luminosity evolution, or a change in $\langle I_e \rangle$.  If we take the value we
measure, \dfp , in Figure \ref{devs} we find \mlfp\ or \dmldz. Our
values are in excellent agreement with the value of $\Delta
\log_{10} M/L_B=-0.415 \pm 0.040$ at $z=0.83$ as found by vv07 but
much smaller than the $d\log_{10} M/L_B = -0.72^{+0.07}_{-0.05} dz$ of
T05.

\subsection{Evolution in the $M/L$ from the Virial Estimator}
\label{mlvar}

\begin{figure*}
\begin{center}
\includegraphics[width=6.4in]{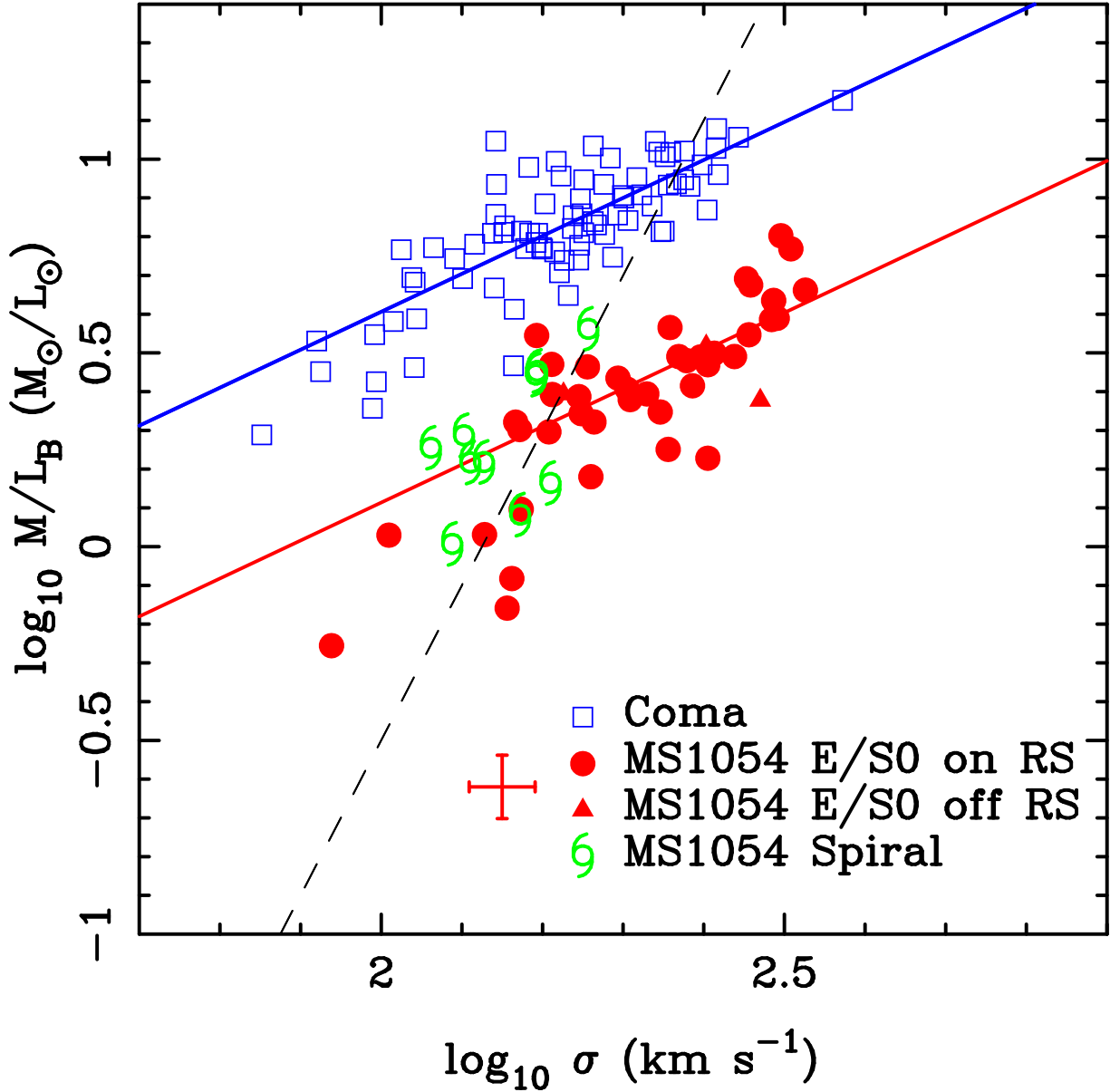}
\end{center}
\caption{ $M/L_B$ as measured by the virial estimator ($M/L_B = 5
  \sigma^2 /(2\pi G r_e \langle I_e \rangle)$) vs. $\sigma$ for the samples of Coma
  and \ms. We use the same symbols as in Figure \ref{fp}. The black
  dashed line is the approximate selection limit for a red-sequence
  galaxy in \ms.  We show the typical errors for \ms\ in the lower
  right.  Our high redshift slope is compatible, within the errors,
  with the slope we find at low redshift.  We find a relation of
  $\log_{10} M/L_B = 0.98 \log_{10}(\sigma/200\ {\rm km\ s^{-1}}) +
  0.90$ for Coma (shown as a blue line) and offset of \mlsig\ for
  red E and S0 galaxies in \ms\ (shown as a red line) and \mlsigt\ for the
  whole sample of $J<21.2$ galaxies, regardless of color or
  morphological type. 
  \label{ml}}

\end{figure*}

We can combine our measurements of the total $B$ light with the $M$ to
estimate a $M/L_B$ value or \[M/L_B = 5 \sigma^2 /(2\pi G r_e \langle I_e \rangle)\]
We show in Figure \ref{ml} the relation between $M/L_B$ and $\sigma$
for Coma and our sample of galaxies in \ms\ from the virial $M/L$
estimator. We fit the results at both low and high redshifts using a
robust linear fitting technique that minimizes the median absolute
deviation and estimate our errors with bootstrapping.  The
bootstrapping is performed in a weighted manner, based on the weights
determined in Section \ref{wgt}.  It appears that the slope has not evolved
between the low and high redshift samples.  In our sample of galaxies
in \ms, we find for the Coma sample, $M/L_B \propto \sigma^{0.98 \pm
  0.10}$, in excellent agreement with the low redshift result of
\citet{vandermarel2007b}, while for our sample in \ms\ we find $M/L_B
\propto \sigma^{1.12 \pm 0.21}$.  If we assume no change in the slope,
we find \mlsigt\ at a fixed $\sigma$ for the whole of the galaxy
population with $J<21.2$.  For the red-sequence, E and S0 population,
we find \mlsig , or no measurable difference.

Our measured $M/L_B$ evolution using the virial estimator is higher
than the \mlfp\ we find in Section \ref{fpevol}.  In principle, these
results should be the same if the evolution in the FP
is caused entirely by $M/L$ evolution.  \citet{vandermarel2007b}
showed that structural evolution, such as caused by higher rotation
rates or size evolution, would cause the disagreement we observe
between the two estimators of $M/L_B$
evolution. \citet{vandermarel2007b} find, for their sample, that the
virial estimator yields a result closer to the $M/L$ as estimate by
the dynamical modeling done by \citet{vandermarel2007b}.  For the rest
of this paper, we will use the virial estimator as a consequence, but
will note how using the FP results changes our results.  This should
avoid some of the bias found in \citet{saglia2010}.  Finally, we add
that \citet{vanderwel2008b} found little evidence for higher rotation
rates in field E and S0 galaxies, thus evolution in rotation alone
does not explain the results of T05 or vdW05.

\subsection{Color Evolution}
\label{colevol}

\begin{figure*}
\begin{center}
\includegraphics[width=6.4in]{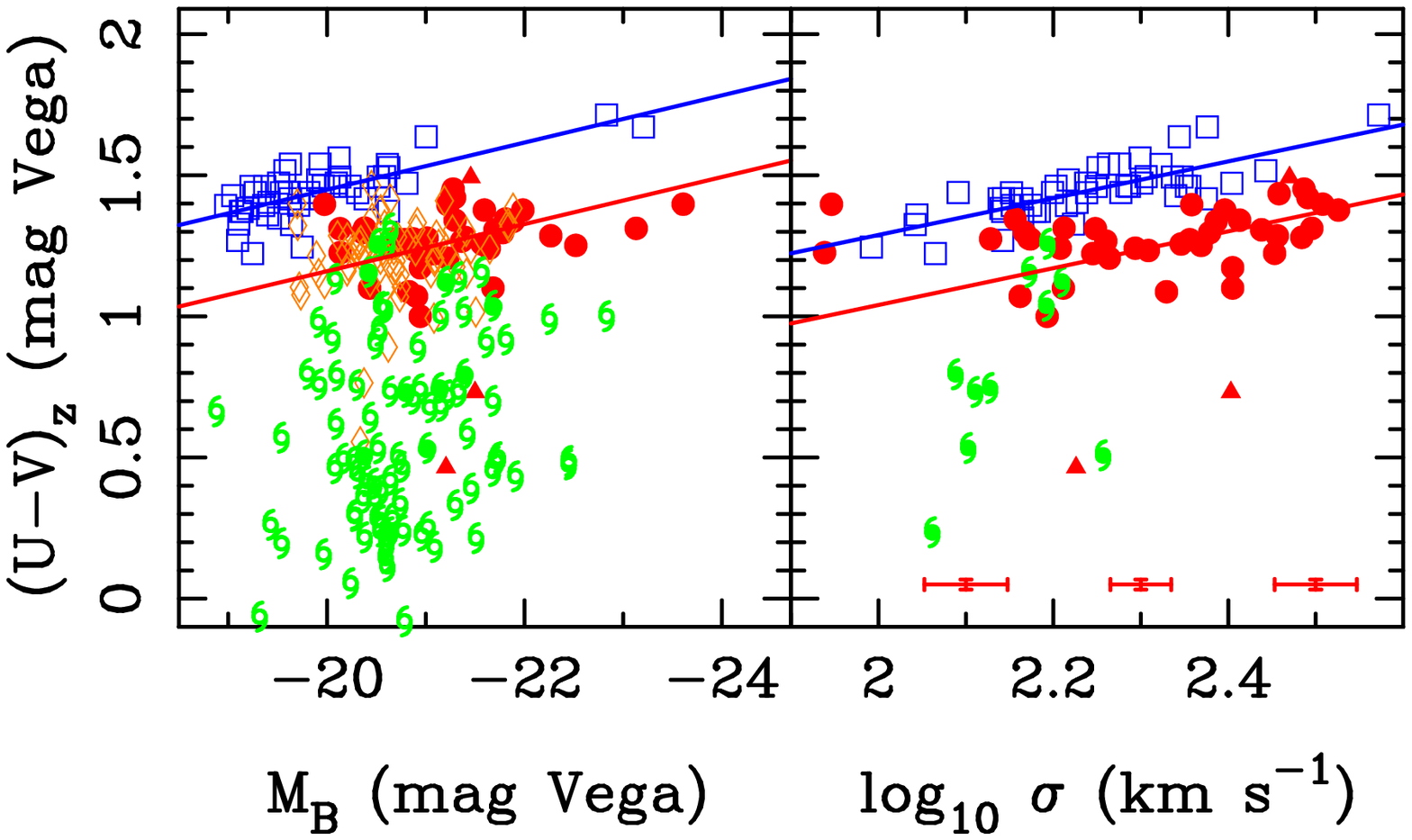}
\end{center}
\caption{ Redshifted color-magnitude and color-$\sigma$ relations. We
  show the $(U-V)_z$ colors both for Coma galaxies ($z=0.0231$) and
  galaxies in \ms\ ($z=0.831$).  We use the same symbols here as in
  Figure \ref{cm}.  The typical error in the color and $\sigma$ is
  shown, as a function of $\sigma$, across the bottom of the right
  panel.  We note that the large $\sigma$ sample is dominated by W04
  which have larger errors in $\sigma$ owing to shorter effective
  exposure times.  We find evolution in the offset of the
  color-magnitude relation and the color-$\sigma$ relation, which
  implies a younger population at higher redshift.  This evolution is
  determined by fitting the color-magnitude and color-$\sigma$
  relation to the Coma galaxies, with the relations shown as blue
  lines, and then computing the best-fitting offset for those galaxies
  on the red-sequence in \ms.  The relations for \ms\ are shown as red
  lines.  We find no statistically significant evidence for a change
  in the slope of either relation when we fit for them to red-sequence
  cluster members. For all red-sequence E and S0 galaxies in our
  sample with $\log_{10}(\sigma) >2.2$, we find \dumv\ of color
  evolution. \label{cmevol} }

\end{figure*}

In Figure \ref{cmevol}, we show that, as expected, the rest-frame
$(U-V)_z$ colors of the \ms\ galaxies are bluer than the colors of the
Coma galaxies. This shift can be seen both at fixed luminosity and at
fixed $\sigma$. Since $\sigma$ is expected to evolve little or not at
all as compared to luminosity, we adopt the color evolution at fixed
$\sigma$ as the quantify of interest.  For Coma, we find the relation
$ (U-V)_z = 0.653 \pm 0.093\ \log_{10} (\sigma/200\ {\rm km\ s^{-1} })
+ 1.475 \pm 0.015 $.

We find no evolution in the slope, as expected from the larger survey
of \citet{mei2009} which found no change in the slope of the
color-magnitude relation for eight $z>0.8$ clusters.  We measure the
offset in the average color for all red-sequence E and S0 galaxies in our
sample with $\log_{10}(\sigma) >2.2$ and we find \dumv\ between
$z=0.831$ and $z=0.023$.  As above, we use bootstrapping to estimate
the errors on the color-$\sigma$ relation in Coma and on the amount of
color evolution.

\section{The  Epoch of Cluster Galaxy Formation }
\label{results}

We use a combination of the $M/L_B$ evolution and the $U-V$ color
evolution to estimate the median epoch of star-formation for galaxies
in our cluster sample.  We estimate the luminosity weighted age of the
stellar populations by computing which single stellar population from
\citet[][hereafter M05]{maraston2005} best reproduces both $M/L_B$ and
$U-V$ evolution we observe.  We will assume that the populations are
coeval and we include all galaxies, regardless of morphological type,
in \ms.  Fitting both the $M/L_B$ evolution and the color-evolution
simultaneously, we find that M05 models with a solar abundance prefer
\mz .  Because of our completeness limits, this is in effect the median age of
formation for galaxies with $\log_{10} \sigma > 2.2$ or, roughly,
a dynamical mass of $M>6\times 10^{10}\ M_{\sun}$.  

If we use the smaller $M/L$ from the offset in the FP, see Section
\ref{fpevol}, we find a higher formation redshift, $z_{\star} =
2.0^{+0.2}_{-0.2}\ $, though well within the range of the errors.  In
fact, the uncertainties in this are as much from the assumptions about
the stellar populations as from the errors on the measurements.  For
example, assuming a metallicity $\log Z/Z_{\sun} = 0.35$  increases the
formation epoch to \hmz .  We note that the pace of $M/L$ evolution and
the pace of the $U-V$ evolution are in mild disagreement with
expectations of a higher metallicity population, but if we assume the
slower pace of evolution from the offset of the FP, we find a better
agreement, though we prefer the estimate from the virial $M/L$
estimator for reasons discussed in Section \ref{mlvar}.

There are significant differences between the models of BC03 and those
of M05, but these are in how the models address the post-main-sequence
evolution, specifically the thermally pulsating asymptotic giant
branch stars.  This can cause significant differences in the near IR
fluxes, but as was shown in \citet{vanderwel2007a} the impact on the
M05 predictions for the optical properties of older stellar
populations is small.  We estimated the typical epoch of formation
for the whole sample using the BC03 models and found $z_{\star} =
1.6^{+0.2}_{-0.1}\ $, in good agreement with the value of \mz\ found
using M05 with the same assumptions.

From the above results we conclude that the epoch of formation
for the galaxies in \ms\ is $z_{\star} = 2.0 \pm 0.3 \pm 0.3\ ({\rm
  sys})$, in good agreement with the results from vv07 for their sample
of higher mass ($>10^{11}\ M_{\sun}$) galaxies.  The systematic error
shows the range of allowed values given both the uncertainties in the
stellar population models as well as the possible different $M/L$ values.

\subsection{Constraints on the Initial Mass Function}
\label{imfmeas}

The evolution of $M/L$ for passively evolving systems depends
critically on the IMF, while the evolution in optical colors depends
less so \citep{tinsley1972}.  This comes about because both the colors
and luminosity of a passively evolving stellar population are
determined by the properties of the stars at the main-sequence
turn-off.  The color evolution depends mostly on the effective
temperature of those stars, and thus is most strongly influenced by
the age and typical metallicity.  The luminosity evolution, however,
depends on not just luminosity of the individual stars, but the
relative numbers of stars as a function of mass.  Thus, our
measurements of the ratio of the pace of color and luminosity
evolution can be used to constrain the IMF of stars at the
main-sequence turn-off, around $0.8-1.0 M_{\sun}$ for old stellar
populations in our target galaxies.

vD08 found that, for cluster galaxies with masses $>10^{11}\
M_{\sun}$, the color and $M/L$ evolution favored an IMF very different
from a Salpeter slope in the mass range of $0.8-1.0 M_{\sun}$.  The
assumptions we made in the previous section use a Salpeter-like IMF to
estimate the formation epoch from the $M/L$ evolution.  vD08, however,
shows that the IMF that most closely agrees with the sample in that
paper yields a formation epoch of $z_{\star} = 3.7^{+2.3}_{-0.8}$ for
cluster galaxies with $>10^{11}\ M_{\sun}$.  In light of the results
of vD08, we have examined our data removing the assumption of a fixed
IMF.  

vD08 provides a useful relation for constraining the IMF with the
evolution in the color and $M/L_B$. vD08 finds that for solar metal
abundances with the M05 models with ages $>$1 Gyr
\[ 2.5 \frac{\Delta \log(M/L_B)}{\Delta (U-V)} = 6.93 - 1.81 x\],
where the terms $\Delta \log_{10} (M/L_B)$ and $\Delta (U-V)$
represent the amount of evolution in the $M/L_B$ and color,
respectively.  The term $x$ is the slope of the IMF, where the
\citet{salpeter1955} value is $x=1.35$.   

\begin{figure*}
\begin{center}
\includegraphics[width=5.2in]{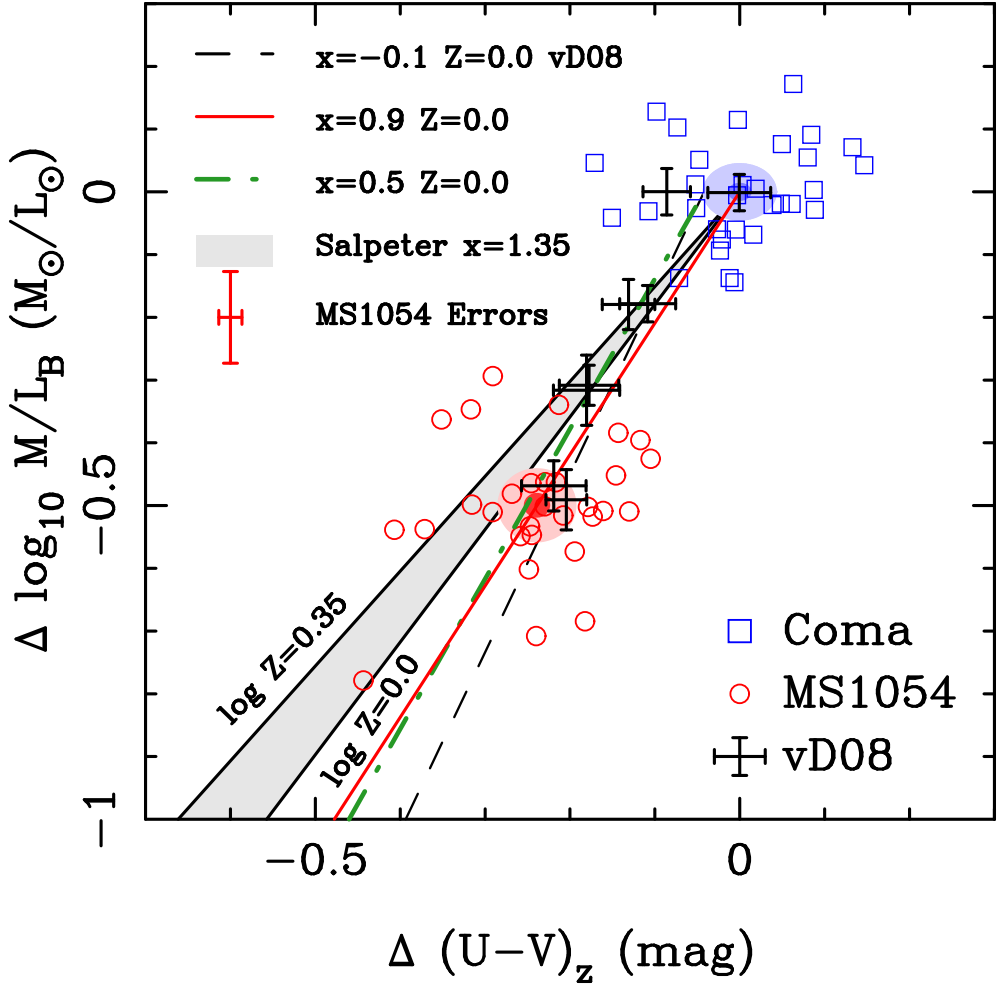}
\end{center}
\caption{ $\Delta \log_{10} M/L_B$ vs. $\Delta(U-V)_z$ for our Coma
  E and S0 sample (blue squares) and for the red-sequence E and S0
  galaxies in \ms\ with $\log_{10} \sigma > 2.2$ (red, open circles).
  For each galaxy, we measure the offset with respect to the Coma
  $M/L_B-\sigma$ (Figure \ref{ml}) and $U-V$-$\sigma$ (Figure
  \ref{cmevol}) relations.  Our mean offsets in $\Delta(U-V) _z$ and
  $\Delta M/L_B$ with 1$\sigma$ error in red and 3 $\sigma$ in
  lighter pink, with our corresponding 1 and 3 $\sigma$ error
  ellipses for Coma in dark and light blue respectively. The typical
  individual errors for \ms\ are shown by a red error-bar in the lower
  right.  We show the expected evolution in $\Delta \log_{10} M/L_B$
  vs. $\Delta(U-V)_z$ for an IMF with a Salpeter slope at $\sim 0.8
  - 1.0 M_{\sun}$ with a metallicity ranging from solar to more than
  twice solar, $\log_{10} (Z/Z_{\sun})= 0.35 $, in the filled gray
  region. The best fitting IMF slope, \imf\, from {\em only} our data
  is shown by the red line. Our sample, alone, shows only tentative
  evidence of a shift in the IMF, at the level of 2.25 standard
  deviations.  The black error bars are the data from vD08, with the
  expected evolution from the best fitting IMF of vD08, $x=-0.1$,
  shown as a dashed line.  Using the lower redshift data set of vD08,
  namely Coma and A2218, we would find \imfvd, shown as a green
  dash-dotted line.  The general agreement with our results (in color)
  and those of vD08 (in black), however, implies that both samples are
  drawn from similar parent samples.  From this, we conclude that the
  stellar populations should have similar IMF slopes. \label{mlbvumv}}
\end{figure*}

We plot $\Delta \log_{10} M/L_B$ and $\Delta(U-V)$ along with the
expected evolution in $\Delta \log_{10} M/L_B$ and $\Delta(U-V)$ in
Figure \ref{mlbvumv}.  We find that our data are in mild disagreement
with the expected evolution from a Salpeter IMF, with a slope of IMF
better described by \imf, shown in Figure \ref{mlbvumv} with a red
line.  This is a difference of 2.25 standard deviations from the Salpeter value
of $x=-1.35$.  This value modestly changes the best fitting formation
epoch from \mz\ to $z_{\star} = 2.0 \pm 0.3$.  As we discuss above,
such a shift is in line with the other systematic uncertainties.

\subsubsection{Contrasting Our Results with vD08}

vD08 found much lower values for $x$, with a preferred value of
$x=-0.1$ for galaxies with a solar metallicity.  We find \imf\ in
large part because we compare only Coma and \ms.  If we include the
lower redshift data from vD08, specifically the data from A2218, we
would find a shallower slope of \imfvd\ illustrated with a green dash-dotted
line in Figure \ref{mlbvumv}.  The key difference between our result
and those of vD08 is not in our high redshift data, but rather the
lower redshift sample.    

vD08's galaxy samples were selected by including all galaxies with
dynamical masses $>10^{11}\ M_{\sun}$.  This is in contrast to our
sample in \ms, which is limited to those galaxies at $\log_{10} \sigma >
2.20$, roughly $6\times 10^{10}\ M_{\sun}$.  If we instead use the
criteria of vD08, namely a subsample of E and S0 galaxies in \ms\ with
$>10^{11}\ M_{\sun}$, we find a $\Delta U-V = 0.22 \pm 0.04$, similar
to the $\Delta U-V = 0.20 \pm 0.04$ found in vD08 and 0.03 mags
smaller than the offset we measure for the full sample of galaxies in
\ms.  So part of the difference in our derived slope of the IMF comes
from our sample probing a larger mass range.  This sample selection
results in a difference in $\Delta U-V = 0.03$ mags or a slope of
$x=0.5 \pm 0.2$ instead of \imf.  The rest of the difference between
our value of \imf\ from our data alone and the $x=-0.1$ as measured by
vD08 is the larger sample of low redshift clusters that vD08 used.  If
we use the same mass selection as vD08, all E and S0 galaxies with
$>10^{11}\ M_{\sun}$ and use the low redshift sample of Coma and
A2218, we recover the $x=-0.1$ found by vD08.

The data we have used in this paper have been analyzed in a consistent
manner throughout.  vD08 used a larger set of data from the literature
that was made as consistent as possible \citep[see][for
details]{vandokkum2006}.  In vD08, it can be seen, both by examining
the equivalent of Figure \ref{mlbvumv} from that paper or from other
measurements of the color and $M/L$ evolution, that the low redshift
data are as important to the conclusions as the high redshift data, a
result reinforced by the change in the slope of the IMF we find when
we include the eight galaxies of A2218.  Because of this, the large
samples of low redshift FP samples, such as \citet{hyde2009b}, should
be used to more fully explore joint evolution in $M/L$ and color.
From this we will have a much more solid measurement of the IMF for
E and S0 systems.

\subsection{Implications from the Lack of Evolution in the Tilt of the
  Fundamental Plane}
\label{slopeage}

We find no evidence for any evolution in the tilt of the FP, nor do we
find evidence for a change in slope of the color-$\sigma$ relation.
Therefore, our data would rule out the scenario where the median
star-formation epoch depends strongly on galaxy mass.  We quantified
this by fitting a model to the color and $M/L$ data in both Coma and
\ms.  This model assumes that the galaxies in Coma and \ms\ is coeval
and have the same metal abundance, and leaves the scatter as a free
parameter at each redshift.  The model parameterizes the formation
epoch as power-law function of the $\sigma$, $t_f = t_o (\sigma/160
{\rm km\ s^{-1}})^{\alpha}$.  Fitting this model to those galaxies
above our magnitude and $\sigma$ limit, we find $\alpha =
0.2^{+0.15}_{-0.1}$ and recover a $t_o$ corresponding to $z_{\star} =
1.5$.  The low value for $\alpha$ means that, in our best-fitting
model, the most massive galaxies ($\log_{10} \sigma = 2.5$) have a
$z_{\star} = 2.3^{+1.3}_{-0.3}$. Our resulting value for $\alpha$ is
on the small side of those reported in the literature, which we will
discuss in later, in Section \ref{compare}.

There is an important caveat to this analysis.  We assume that the
whole population in both clusters are coeval.  Because the cluster
population grows by accretion of galaxies from outside of the cluster,
however, the population of galaxies in \ms\ represents only a subset
of the $z=0$ cluster galaxy population.  Thus, we would overestimate
the epoch of cluster galaxy formation as we would only include those
galaxies that are already E, S0 or early-type spirals at $z=0.83$. If
this process is dramatic, we would expect to find rapid evolution in
the fraction E and S0 galaxies in clusters.  However, for mass
selected samples, the amount of observed evolution in the fraction of
E and S0 galaxies is minimal \citep{holden2006,holden2007}.
\citet{pvd_mf2001} provide a method for calculating the impact of this
bias.  We find if we assume that 90\% of the galaxies were in our
sample at $z=1$, there is no shift in the epoch of formation.
Assuming that fraction is only 50\%, the effective epoch of formation
shifts downward by $\Delta z = 0.2$, $z_{\star} = 1.4$ for galaxies
with $\log \sigma = 2.2$.  It is likely that this fraction actually
varies with velocity dispersion, so, in effect, the actual slope of
the relation between formation epoch and dispersion is mildly steeper
than we measure.  If we assume that 50\% of the $\log \sigma =2.2$
galaxies were not present in the sample while 90\% of the $\log \sigma
=2.5$ were present, this shifts the resulting $\alpha$ from $\alpha =
0.2^{+0.15}_{-0.1}$ to $\alpha = 0.25^{+0.15}_{-0.1}$ , a very modest
change.

\subsubsection{Implications for the IMF }

\citet{renzini2006} shows that a trend in the IMF with mass would
cause the tilt in the FP to evolve with redshift.  Because our
measured slope of the FP changes little, if at all, we can immediately
conclude that the IMF of the lower mass galaxies in our sample is
similar to that of the higher mass systems.  In Section \ref{imfmeas}, we
show that most of the difference between the slope of the IMF from our
sample, $x=0.9$, and the slope of the IMF as determined by vD08,
$x=-0.1$, comes from the larger sample of clusters used in vD08.  The
remaining difference, a change in the slope $\Delta x = 0.4$,
comes about from our sample of lower mass galaxies in \ms.  Thus, the
IMF for cluster galaxies has, at most, a mild mass dependence for
galaxies with $\log_{10} \sigma > 2.2$, and a mass dependence that
appears in agreement with the results of \citet{treu2010}.

As many lower mass cluster galaxies must have formed in more
field-like environments before falling into the cluster
\citep[eg.][and references therein]{patel2009a,berrier2009}, then the
IMF for many field E and S0 galaxies should be similar to what is found
for cluster galaxies at similar masses.  This follows regardless of
whether the conclusions of vD08 are correct or, in contrast, the
conclusions of \citet{treu2010} are correct.  Therefore, if there is a
different IMF for E and S0 galaxies, it should be prevalent in all
moderate to high mass galaxies regardless of environment.

\subsection{Comparison with Color Measurements}

A number of color-based measurements of the formation epoch of cluster
galaxies have been made.  B06 found that the scatter in the colors in
\ms\ and another $z=0.83$ cluster, \cl, implied a star formation epoch
of $z_{\star} \simeq 2.2$, in good agreement with our results.
\citet{mei2009} found a similar scatter in the colors of the E and S0
population in a sample of eight clusters, including the two in B06,
indicating a similar formation epoch for the $z\sim1$ cluster
population. \citet{holden2004} measured $z_{\star} = 3^{+2}_{-1}$ by
fitting the evolution of the zero-point of the color-magnitude
relation of E and S0 galaxies, as opposed to the scatter as was done
in B06, for a sample of 24 clusters spanning a redshift range of $0.3
< z < 1.3$.  \citet{eisenhardt2008} and \citet{mancone2010} find a
similar formation epoch for the galaxies on the red-sequence in a
much larger sample of $z>1$ clusters than in \citet{holden2004}.  All
of these results all point to $z=2-3$ being the important epoch for
the star formation of cluster early-type systems.  Interestingly,
\citet{eisenhardt2008} find evidence that the average formation epoch
of the cluster red-sequence increases for the higher redshift cluster
galaxies, the expected result if more galaxies are joining the red-sequence
over time.

At $z\sim2-3$, \citet{brammer2007}, \citet{zirm2008},
\citet{kriek2008a} and \citet{brammer2009} all find a well-defined
red-sequence in both clusters and the field environment. These
early-type systems have finished forming their stars at $z\sim3$, a
time higher than the mean age as measured by colors or by the
FP. However, these galaxies represent only the oldest 10\% of the
early-type population, and so are consistent with a more typical epoch
of $z_{\star} \sim 2$ for the majority of the population.  Therefore, color measurements
show that $\sim L^{\star}$ galaxies have stellar populations with
formation times of $z_{\star}\sim2-3$, an epoch when a sizable fraction of
the stars that exist today were formed and entirely consistent with
our FP measurements.



\subsection{Comparison with Archaeological Studies of Present-day
  Galaxies}
\label{compare}

\citet{trager2000a,trager2000b} found that age and dispersion were
part of a hyper-plane also including metallicity and $\alpha$
enhancement.  In general, higher dispersion galaxies have older
population ages at a fixed metallicity and galaxies in denser
environments show older ages.  A number of papers have built on these
results, finding a strong correlation between the properties of 
stellar populations and velocity dispersions of galaxies for very
large samples
\citep{nelan2005,gallazzi2006,graves2009a,graves2009b,smith2009}. Interestingly
dynamical mass is not the important variable, rather
\citet{graves2009a,graves2009b} and \citet{smith2009} both found that
$\sigma$ is the driving parameter for determining the stellar
populations of $z=0$ early-type galaxies.

The ages derived from these measurements are strikingly young for lower
dispersion galaxies, roughly 30\%-50\% of galaxies at $\log_{10} \sigma =2.2$
having formation ages corresponding to $z\sim0.8$. \citet{trager2008}
found similar results for E and S0 galaxies in the Coma cluster, a typical
age of 5-8 Gyr, or roughly a $z_{\star}\sim 0.5-1$, regardless of
galaxy mass. Correspondingly, many authors find a steep relation
between the age of the stellar population and velocity dispersion,
with logarithmic slopes of $0.4-0.6$ common
\citep{trager2000b,nelan2005,smith2009} and values up to $\sim1.1$
found \citep{bernardi2003d}.  \citet{kelson2006} found a strong
relation between metallicity and $\sigma$, but, in contrast, that
after accounting for the selection limits the population of \clt\
showed no age variation with $\sigma$.

Part of the reason for the difference in our results, a shallow
age-dispersion relation, and the steeper ones found by others comes
about from our use of broad band photometric data as compared to
fitting models of stellar populations to spectra.  \citet{tortora2009}
found that the slope of the age-$\sigma$ relation is much shallower
when stellar populations are fit to only broad band photometry.
Further, \citet{cooper2009} found that galaxies in high density
environments with the same broad band color and magnitude as galaxies
in low density environments, had both larger ages and larger
metallicities using models fit to spectra from \citet{gallazzi2006}.
This shows that the population parameters derived from spectra are not
the same as those derived from color measurements alone.
\citet{trager2009} provides a potential framework to interpret these
seemingly contradictory results.  \citet{trager2009} find that the
ages that result from fitting simple stellar population models to
spectral features are likely when the last 5\%-10\% of the stars
formed.  Thus these young apparent ages are not when the median star
was formed, but rather the epoch when the last star formation
occurred.

\subsection{The Field versus the Cluster Early-type Population}

Our sample consists entirely of cluster galaxies, so part of the
reason we find a different result from field samples such as T05 or
vdW05 could simply be environment.  \citet{bernardi2003c},
\citet{bernardi2006}, \citet{cooper2009}, \citet{labarbera2010b} and \citet{saglia2010} all
find differences in the FP or the stellar populations of galaxies such
that galaxies in clusters appear older than the same mass field
galaxies.

Interestingly, T05 show, in Figure 14 of their paper, that most
massive ($>10^{11}\ M_{\sun}$) galaxies have a formation epoch of
$z\sim 2-3$.  vv07 perform a separate analysis and find a similar
result.  The main differences between our results and T05 are at
$<10^{11}\ M_{\sun}$.  We find that $\sim L^{\star}$ cluster galaxies
have a typical formation epoch of $z_{\star} =1.7$ to $z_{\star}=2$,
whereas T05 find $z_{\star} \sim 1.2$ for similar mass galaxies in
their sample.  In our \ms\ sample, we have galaxies that show rapid
enough $M/L$ evolution to be consistent with a formation epoch of
$z_{\star} \sim 1.2$, but the typical low $\sigma$ galaxy has a much
higher $M/L$ (see Figure \ref{ml}).  In the context of our model
presented in the previous section where $t_{\star} \propto
(\sigma)^{\alpha}$, the results of T05 prefer an $\alpha \sim
0.4-0.45$ as opposed to our $\alpha \sim 0.2$.  There are two possible
explanations for why our results differ from T05.  First, the selection
limits of T05 $z$ band filter galaxy selection, as opposed to our
redder $J$, biased the resulting sample towards bluer galaxies than
our sample,. The second possibility is that the results of T05 are
different for lower mass field galaxies because lower mass field
galaxies have a different star-formation history as compared with
lower mass cluster galaxies.  \citet{patel2009a} and
\citet{patel2009b} do find that field and cluster galaxies have
different star-formation rates and color distributions at $z=0.83$,
possibly explaining the difference between our results and T05.

T05 explain the apparently young stellar populations seen in lower
mass galaxies because of small recent episodes of star-formation and
not because the galaxy population is, as a whole, young, akin to the
model of \citet{trager2009}.  This picture is given more support when
compared with the recent results of \citet{thomas2009}, which find
that a small fraction of galaxies in the local universe with an E and
S0 morphology appear to have had recent, small episodes of
star-formation.  Labeling these young galaxies as ``rejuvenated'',
\citet{thomas2009} found the fraction of these galaxies increases at
lower velocity dispersions, mimicking the steep age-dispersion
relation found in T05 and in other work.  \citet{rogers2010} also
found a higher fraction of galaxies with recent star-formation in
lower mass dark matter halos.  All of this is consistent with the
\citet{trager2009} picture.  To explain the results of, say,
\citet{cooper2009} or \citet{rogers2010}, these events would be more
common in field like environments.  These are likely the handful of
galaxies we observe with bluer colors and lower $M/L_B$ values at
lower values of $\sigma$.  This picture provides a natural explanation
for the build up of the red-sequence from $z\sim1$ until today, see,
for example, \citet{harker2006} or \citet{ruhland2009}, and could
provide the progenitors for the apparently young Coma galaxies in
\citet{trager2008}.

\section{Summary }
\label{summary}

We have compiled a sample of velocity dispersion measurements of
cluster galaxies spanning a broad range in mass at $z=0.831$.  Our
sample goes much farther down the mass function than many previous
cluster and field samples.  Our sample was selected only by $J$
magnitude, effectively in rest-frame $r$ at this redshift, closely
mimicking the selection used for many low-redshift samples.  No
morphological or color information were used in the selection.  With
{\em HST}/ACS imaging in the bandpass closest to the rest-frame $B$,
we determined the size and surface brightnesses of the galaxies in our
sample.  We found that our success at measuring velocity dispersions
depended mostly on the magnitude of the galaxies, with no apparent
bias in size or surface brightness.  We did find that, for the
early-type spirals in our sample,  we are less likely to measure
the dispersions of galaxies with bluer colors, so we proceeded by
giving those galaxies higher weight in our analysis.

We determined from our sample the best fitting FP \(
\log r_e = a \log \sigma + b \log \langle I_e \rangle + c\), and measure the color and
$M/L$ evolution of the early-type population of sub-$L^{\star}$
galaxies.  From our analysis, we conclude the following.

\begin{itemize}

\item We find no evidence that the slope of the FP
  evolves between Coma and the $z=0.83$ cluster \ms.  Our sample
  containing galaxies with $\sigma \sim 160$ km s$^{-1}$ have a 
  FP with the same slope as is found at $z=0$.

\item The evolution in $M/L$ is observed to be similar to that seen by
  some previous authors (W04; Holden \etal\ 2005; vv07), where the
  $M/L$ ratio of the cluster population evolves with \mlsig\
  from $z=0.023$ and $z=0.83$, or \dmldz, with only a mild $\sigma$
  dependence on the amount of $M/L$ evolution.  This in marked
  contrast to more rapid evolution in field studies, such as T05 or
  vdW05, or some other cluster results such as \citet{jorgensen2006}.

\item Using the evolution in the $M/L_B$ and in the colors of the
  galaxies, we find a formation epoch of \mz\ to \hmz, the range
  depending on the stellar population assumptions we make.  This is
  similar to previous measurements, see for example
  \citet{kelson2000c,wuyts2004,holden2005}, and vv07 and is in good
  agreement with color estimates such as B06 and \citet{mei2009}.

\item The lack of evolution in the tilt of the FP and in the
  color-$\sigma$ relations implies little trend in the formation
  epoch of cluster galaxies with $\sigma$.  We find $z_{\star} =
  2.3^{+1.3}_{-0.3}$ for the most massive, $>300\ {\rm km\ s^{-1}}$,
  cluster galaxies, while we find $z_{\star} = 1.7^{+0.3}_{-0.2}$ for
  those with $\sim160\ {\rm km\ s^{-1}}$.  This same lack of evolution
  in the tilt of the FP and color-$\sigma$ relation also implies a
  similar IMF for the lower mass cluster galaxies as is seen at higher
  masses, with a upper limit of $\Delta x \sim 0.4$ over the mass
  range we cover, where $x$ is the slope of the IMF.

\end{itemize}

\vspace{.3cm}

The first author thanks Hans-Walter Rix and the rest of
the MPIA for their hospitality during part of the writing of this
paper  and thanks Pieter van Dokkum for useful
discussions.  The authors particularly thanks the referee
for a large number of helpful suggestions that significantly improved
the paper and its presentation.  The authors wish to recognize and
acknowledge the very significant cultural role and reverence that the
summit of Mauna Kea has always had within the indigenous Hawaiian
community.  We are most fortunate to have the opportunity to conduct
observations from this mountain.This research has made use of the
NASA/IPAC Extragalactic Database (NED) which is operated by the Jet
Propulsion Laboratory, California Institute of Technology, under
contract with the National Aeronautics and Space Administration.  This
research was funded, in part, by grants HST-GO-09772.01-A.

\appendix
\LongTables

\section{\ms\ Sample }
\label{msdata}

\subsection{Fitting Surface Brightness Profiles}

For each galaxy, we fit the surface brightness profiles in multiple
steps.  A S{\'e}rsic model was fit first, using values from a catalog
derived by {\tt SExtractor} as initial guesses but with a restriction
on $n$ of $1 \le n \le 4$.  The output of the S{\'e}rsic model was
used an initial guess for the de Vaucouleur's model fit, and the
S{\'e}rsic model was refit using the previous iteration as an initial
guess.  If the best fitting S{\'e}rsic index is $n=1$, we refit
freezing $n=1$.  This was done to ensure convergence.

Galaxies in crowded environment, generally those near the center of
the cluster, were fit simultaneously along with all of those galaxies'
neighbors.  We use the sizes from the SExtractor catalog to find
galaxies with close neighbors.  All galaxies within a radius twice the
size of a target galaxy were fit along with the target galaxy.  In
addition, galaxies outside of that radius but with sizes such that the
target galaxy lies within twice their size were also included in the
joint fit.  This later criterion was chosen because particularly large
galaxies, like the brightest galaxy of the cluster, often have extended surface
brightness profiles.  After the fitting process, if the radii of the
target galaxy were significantly larger than the initial {\tt
  SExtractor} guess, we refit including more neighboring galaxies.  In
all cases, neighboring galaxies were fit with a S{\'e}rsic model
constrained to lie in $1 \le n \le 4$.

We note here that the ACS magnitudes and surface brightnesses have
been adjusted by the offsets as we computed from simulations that are
listed in Section \ref{measure}.

\subsection{Errors on the Morphologies}

A number of galaxies in our sample were morphologically typed multiple
times.  All of the galaxies in our sample were classified by Marc
Postman twice, once for \citet{postman2005} and once for the parent
sample that the galaxies in this paper are drawn from.  In addition,
many of these galaxies were classified in WFPC2 imaging
\citep{tran2007}.  For the purposes of this paper, the main
classification is the distinction between a late-type or an early-type
system, where the former are spiral and irregular galaxies while the
later are ellipticals and S0 galaxies.  The comparison in
\citet{postman2005} finds that 10\% of the galaxies in the ACS sample
are identified as spirals in the WFPC2 or vice-versa.  When comparing
that two separate classifications done by Marc Postman, we find that
92\% of galaxies are grouped in the same two broad bins.

\subsection{Summary of the Data}

We list all of the galaxies targeted in Table \ref{ms1054data}.
For each galaxy, we list the redshift and dispersion, if available.
For a summary of why some galaxies do not have dispersions, see Section
\ref{veldisp}.  The galaxies in the sample of W04 are listed with a
``w'' in front of the identification number.  The galaxies with
redshifts are those we re-observed as part of our program.  All of
the other galaxies have spectroscopic information from W04 which does
not list redshifts, though those can be found in \citet{tran2007}.

\begin{deluxetable*}{lllllllllllll}
\tablecaption{Data for Galaxies in \ms \label{ms1054data}}
\tablecolumns{13}
\tabletypesize{\footnotesize}
\tablehead{
\colhead{ID} & \colhead{$z$} & \colhead{$\log_{10} \sigma$}  &  \colhead{$\log_{10} r_e$\tablenotemark{a}} & 
\colhead{$q$\tablenotemark{a}} & \colhead{$n$ \tablenotemark{a}} & \colhead{$-0.4 \mu_B$\tablenotemark{a}}  &  \colhead{$B_z$} & \colhead{$(U-V)_z$\tablenotemark{b}} & 
\colhead{$r_z$} &  \colhead{T} & \colhead{$J$}  & \colhead{$I-J$} \\
\colhead{} & \colhead{} & \colhead{$(\log_{10}{\rm km\ s^{-1}})$ }  &  \colhead{($\log_{10}{\rm kpc}$)} &
\colhead{} & \colhead{} & \colhead{(-0.4 mag)}  &  \colhead{(mag)} & \colhead{(mag)} & 
\colhead{(mag)} & \colhead{} & \colhead{(mag)}  & \colhead{(mag)} \\
}
\startdata
3500 & 0.836 & ... & 0.679 $\pm$ 0.017 & 0.80 & 1.6 & -9.04 & 22.85 & 0.04 & 22.35 & 6 & 20.75 & 0.52 \\ 
3533 & 0.837 & 2.294 $\pm$ 0.028 & 0.630 $\pm$ 0.009 & 0.84 & 4.0 & -8.59 & 21.97 & 1.24 & 21.24 & -5 & 19.56 & 1.15 \\ 
4336 & 0.835 & 2.227 $\pm$ 0.044 & 0.200 $\pm$ 0.042 & 0.64 & 3.5 & -8.48 & 23.85 & 1.14 & 23.25 & -2 & 21.61 & 0.89 \\ 
4685 & 0.835 & ... & 0.478 $\pm$ 0.015 & 0.47 & 1.0 & -8.91 & 23.53 & 0.38 & 23.07 & 3 & 21.48 & 0.46 \\ 
4846 & 0.697 & ... & 0.492 $\pm$ 0.036 & 0.90 & 3.7 & -8.73 & 23.02 & 0.35 & 21.71 & -5 & 20.05 & 1.01 \\ 
4928 & 0.851 & ... & 0.542 $\pm$ 0.015 & 0.40 & 1.0 & -8.96 & 23.35 & 0.50 & 22.31 & 4 & 20.64 & 1.09 \\ 
5108 & 0.840 & 2.208 $\pm$ 0.030 & 0.287 $\pm$ 0.010 & 0.71 & 4.0 & -8.29 & 22.95 & 1.24 & 22.19 & -2 & 20.52 & 1.13 \\ 
5152 & 0.837 & ... & 0.165 $\pm$ 0.016 & 0.92 & 4.0 & -8.17 & 23.26 & 1.11 & 22.79 & -5 & 21.15 & 0.85 \\ 
5234 & 0.837 & 1.938 $\pm$ 0.066 & 0.046 $\pm$ 0.012 & 0.93 & 4.0 & -8.02 & 23.48 & 1.23 & 22.63 & -5 & 20.98 & 0.97 \\ 
5588 & 0.824 & ... & 0.569 $\pm$ 0.007 & 0.92 & 1.0 & -8.79 & 22.77 & 0.37 & 22.18 & 4 & 20.52 & 1.04 \\ 
5795 & 0.842 & 2.245 $\pm$ 0.031 & 0.474 $\pm$ 0.009 & 0.75 & 4.0 & -8.50 & 22.52 & 1.22 & 21.42 & -2 & 19.73 & 1.25 \\ 
5894 & 0.827 & ... & 0.297 $\pm$ 0.023 & 0.39 & 1.0 & -8.79 & 24.13 & 1.35 & 1.64 & 1 & 0.10 & 0.00 \\ 
5987 & 0.829 & 2.062 $\pm$ 0.046 & 0.526 $\pm$ 0.004 & 0.79 & 1.0 & -8.78 & 22.97 & 0.23 & 22.53 & 6 & 20.91 & 0.65 \\ 
6169 & 0.777 & 1.946 $\pm$ 0.072 & 0.028 $\pm$ 0.016 & 0.81 & 2.7 & -8.05 & 23.63 & 1.40 & 22.68 & -5 & 21.01 & 1.12 \\ 
6191 & 0.824 & 2.356 $\pm$ 0.028 & -0.004 $\pm$ 0.013 & 0.45 & 4.2 & -7.66 & 22.82 & 1.27 & 22.10 & -1 & 20.43 & 1.07 \\ 
6333 & 0.837 & 2.212 $\pm$ 0.028 & 0.264 $\pm$ 0.013 & 0.41 & 2.0 & -8.36 & 23.23 & 1.31 & 22.62 & -2 & 20.93 & 1.24 \\ 
6410 & 0.833 & 2.305 $\pm$ 0.030 & -0.004 $\pm$ 0.013 & 0.49 & 3.6 & -7.92 & 23.47 & 1.20 & 22.92 & -2 & 21.25 & 1.14 \\ 
6426 & 0.832 & ... & 0.097 $\pm$ 0.021 & 0.72 & 3.0 & -8.38 & 24.11 & 1.22 & 23.19 & -2 & 21.50 & 1.27 \\ 
6448 & 0.849 & 2.111 $\pm$ 0.063 & 0.456 $\pm$ 0.007 & 0.54 & 1.4 & -8.58 & 22.81 & 0.73 & 22.52 & 1 & 20.91 & 0.64 \\ 
6594 & 0.832 & 2.264 $\pm$ 0.029 & 0.426 $\pm$ 0.016 & 0.72 & 3.9 & -8.35 & 22.39 & 1.21 & 21.66 & -4 & 19.98 & 1.16 \\ 
6630 & 0.842 & ... & -0.004 $\pm$ 0.010 & 0.60 & 1.0 & -8.15 & 24.04 & 0.63 & 23.04 & 0 & 21.39 & 0.94 \\ 
6842 & 0.830 & 2.010 $\pm$ 0.041 & 0.293 $\pm$ 0.020 & 0.71 & 3.9 & -8.43 & 23.26 & 1.04 & 22.86 & -4 & 21.21 & 0.96 \\ 
7075 & 0.831 & 2.130 $\pm$ 0.087 & 0.040 $\pm$ 0.018 & 0.79 & 2.5 & -8.22 & 24.01 & 1.16 & 23.43 & -9 & 21.76 & 1.11 \\ 
7083 & 0.835 & 2.167 $\pm$ 0.055 & 0.275 $\pm$ 0.012 & 0.77 & 2.6 & -8.39 & 23.25 & 1.30 & 21.58 & -4 & 19.88 & 1.40 \\ 
7088 & 0.834 & ... & 0.218 $\pm$ 0.014 & 0.80 & 4.0 & -8.49 & 23.77 & 1.29 & 22.48 & -5 & 20.79 & 1.31 \\ 
7247 & 0.821 & 2.193 $\pm$ 0.030 & 0.680 $\pm$ 0.030 & 0.48 & 3.5 & -8.97 & 22.66 & 1.00 & 22.25 & -1 & 20.59 & 1.02 \\ 
7415 & 0.831 & 2.309 $\pm$ 0.028 & 0.264 $\pm$ 0.009 & 0.52 & 2.1 & -8.15 & 22.71 & 1.23 & 21.67 & -2 & 20.00 & 1.19 \\ 
7534 & 0.839 & 2.257 $\pm$ 0.041 & 0.337 $\pm$ 0.008 & 0.45 & 1.0 & -8.51 & 23.24 & 0.50 & 22.28 & 4 & 20.62 & 1.01 \\ 
7613 & 0.830 & 2.103 $\pm$ 0.041 & 0.628 $\pm$ 0.003 & 0.70 & 1.0 & -8.83 & 22.59 & 0.53 & 22.25 & 8 & 20.62 & 0.72 \\ 
7648 & 0.833 & 2.248 $\pm$ 0.032 & 0.043 $\pm$ 0.018 & 0.24 & 1.7 & -8.02 & 23.48 & 1.31 & 22.43 & -1 & 20.74 & 1.26 \\ 
7778 & 0.826 & 2.211 $\pm$ 0.028 & 0.363 $\pm$ 0.013 & 0.48 & 2.1 & -8.54 & 23.18 & 1.10 & 22.11 & -1 & 20.45 & 1.04 \\ 
7937 & 0.848 & 2.172 $\pm$ 0.033 & 0.046 $\pm$ 0.018 & 0.28 & 2.4 & -7.91 & 23.19 & 1.15 & 22.30 & 1 & 20.62 & 1.20 \\ 
8438 & 0.831 & 2.210 $\pm$ 0.033 & 0.366 $\pm$ 0.016 & 0.34 & 2.6 & -8.23 & 22.40 & 1.12 & 21.67 & 1 & 20.01 & 1.11 \\ 
8572 & 0.825 & 2.162 $\pm$ 0.027 & 0.099 $\pm$ 0.005 & 0.82 & 4.0 & -7.82 & 22.70 & 1.07 & 22.03 & -5 & 20.37 & 0.99 \\ 
8740 & 0.825 & 2.487 $\pm$ 0.025 & 0.316 $\pm$ 0.014 & 0.48 & 3.4 & -8.11 & 22.33 & 1.45 & 21.18 & -1 & 19.50 & 1.20 \\ 
8771 & 0.831 & 2.088 $\pm$ 0.052 & 0.509 $\pm$ 0.010 & 0.75 & 4.0 & -8.45 & 22.22 & 0.79 & 21.67 & 1 & 20.01 & 0.96 \\ 
8801 & 0.827 & 2.226 $\pm$ 0.027 & 0.568 $\pm$ 0.007 & 0.84 & 4.0 & -8.64 & 22.40 & 0.47 & 21.60 & -2 & 19.92 & 1.16 \\ 
8839 & 0.841 & ... & 0.298 $\pm$ 0.010 & 0.68 & 1.4 & -8.48 & 23.36 & 0.12 & 22.84 & 8 & 21.24 & 0.58 \\ 
9061 & 0.838 & 2.174 $\pm$ 0.029 & 0.174 $\pm$ 0.007 & 0.80 & 4.0 & -8.05 & 22.90 & 1.27 & 22.12 & -4 & 20.43 & 1.31 \\ 
9145 & 0.835 & 2.172 $\pm$ 0.111 & 0.489 $\pm$ 0.012 & 0.85 & 4.0 & -8.56 & 22.59 & 1.28 & 21.53 & -5 & 19.88 & 1.01 \\ 
9288 & 0.836 & 2.127 $\pm$ 0.034 & 0.542 $\pm$ 0.012 & 0.89 & 2.0 & -8.61 & 22.47 & 0.74 & 21.70 & 4 & 20.04 & 1.03 \\ 
9306 & 0.829 & 2.358 $\pm$ 0.027 & 0.464 $\pm$ 0.010 & 0.75 & 4.0 & -8.42 & 22.38 & 1.40 & 21.42 & -4 & 19.75 & 1.08 \\ 
10126 & 0.833 & ... & 0.383 $\pm$ 0.005 & 0.93 & 1.0 & -8.52 & 23.03 & -0.04 & 22.83 & 8 & 21.24 & 0.46 \\ 
10441 & 0.829 & 2.193 $\pm$ 0.036 & 0.394 $\pm$ 0.011 & 0.37 & 1.0 & -8.57 & 23.10 & 1.25 & 22.20 & 1 & 20.50 & 1.29 \\ 
10480 & 0.660 & ... & 0.337 $\pm$ 0.043 & 0.37 & 1.3 & -8.97 & 24.38 & 0.03 & 23.21 & 1 & 21.61 & 0.55 \\ 
10806 & 0.827 & ... & 0.673 $\pm$ 0.012 & 0.82 & 2.1 & -8.75 & 22.15 & 0.37 & 21.79 & 4 & 20.16 & 0.78 \\ 
10829 & 0.857 & ... & 0.633 $\pm$ 0.013 & 0.64 & 1.0 & -9.09 & 23.22 & 0.38 & 22.55 & 6 & 20.91 & 0.85 \\ 
11217 & 0.828 & 2.256 $\pm$ 0.035 & 0.002 $\pm$ 0.013 & 0.64 & 2.1 & -8.06 & 23.80 & 1.01 & 22.99 & -5 & 21.33 & 1.03 \\ 
11236 & 0.828 & ... & 0.524 $\pm$ 0.016 & 0.30 & 1.0 & -8.86 & 23.19 & 0.49 & 22.69 & 3 & 21.09 & 0.53 \\ 
11297 & 0.836 & 2.128 $\pm$ 0.027 & 0.249 $\pm$ 0.013 & 0.92 & 2.9 & -8.13 & 22.73 & 1.27 & 22.04 & -2 & 20.39 & 0.97 \\ 
11461 & 0.822 & ... & 0.638 $\pm$ 0.011 & 0.61 & 3.6 & -8.32 & 21.26 & 0.94 & 20.63 & 4 & 18.98 & 0.96 \\ 
11558 & 0.830 & 2.192 $\pm$ 0.029 & 0.855 $\pm$ 0.034 & 0.42 & 4.0 & -9.02 & 21.93 & 1.03 & 21.07 & 4 & 19.38 & 1.36 \\ 
11870 & 0.827 & ... & 0.334 $\pm$ 0.023 & 0.47 & 1.1 & -8.81 & 24.00 & 0.47 & 22.99 & 0 & 21.33 & 1.03 \\ 
11918 & 0.833 & ... & -0.319 $\pm$ 0.021 & 0.75 & 4.0 & -6.91 & 22.52 & -0.35 & 21.76 & 1 & 20.14 & 0.68 \\ 
11945 & 0.838 & 2.329 $\pm$ 0.030 & 0.210 $\pm$ 0.008 & 0.56 & 4.0 & -8.07 & 22.78 & 1.09 & 22.07 & -1 & 20.39 & 1.22 \\ 
w1192 & 0.840 & 2.156 $\pm$ 0.026 & 0.395 $\pm$ 0.005 & 0.75 & 4.0 & -8.05 & 21.80 & 1.34 & 20.70 & -4 & 19.02 & 1.15 \\ 
w1649 & ... & 2.386 $\pm$ 0.050 & 0.541 $\pm$ 0.006 & 0.78 & 4.0 & -8.31 & 21.72 & 1.34 & 20.94 & -4 & 19.26 & 1.24 \\ 
w2409 & ... & 2.458 $\pm$ 0.050 & 0.408 $\pm$ 0.009 & 0.68 & 3.2 & -8.29 & 22.34 & 1.43 & 21.56 & -4 & 19.87 & 1.26 \\ 
w3058 & 0.832 & 2.496 $\pm$ 0.028 & 1.207 $\pm$ 0.010 & 0.72 & 3.7 & -9.15 & 20.48 & 1.31 & 19.96 & -5 & 18.27 & 1.28 \\ 
w3768 & ... & 2.346 $\pm$ 0.047 & 0.427 $\pm$ 0.004 & 0.90 & 4.0 & -8.21 & 22.04 & 1.26 & 21.43 & -5 & 19.74 & 1.30 \\ 
w3910 & ... & 2.470 $\pm$ 0.062 & 0.163 $\pm$ 0.007 & 0.42 & 4.0 & -7.73 & 22.15 & 1.49 & 21.58 & -2 & 19.91 & 1.18 \\ 
w4345 & ... & 2.526 $\pm$ 0.044 & 0.546 $\pm$ 0.009 & 0.68 & 3.4 & -8.28 & 21.62 & 1.38 & 20.74 & -5 & 19.05 & 1.28 \\ 
w4520 & ... & 2.508 $\pm$ 0.040 & 1.339 $\pm$ 0.013 & 0.64 & 4.9 & -9.22 & 20.00 & 1.40 & 19.72 & -5 & 18.03 & 1.34 \\ 
w4705 & ... & 2.403 $\pm$ 0.062 & 0.455 $\pm$ 0.007 & 0.28 & 1.6 & -8.29 & 22.11 & 0.73 & 21.15 & -1 & 19.51 & 0.89 \\ 
w4926 & ... & 2.491 $\pm$ 0.053 & 0.267 $\pm$ 0.007 & 0.46 & 4.0 & -8.00 & 22.32 & 1.42 & 21.48 & -3 & 19.78 & 1.35 \\ 
w5280 & ... & 2.413 $\pm$ 0.052 & 0.335 $\pm$ 0.011 & 0.36 & 2.3 & -8.14 & 22.31 & 1.34 & 21.42 & -2 & 19.73 & 1.23 \\ 
w5298 & ... & 2.453 $\pm$ 0.060 & 0.331 $\pm$ 0.012 & 0.51 & 2.5 & -8.24 & 22.60 & 1.22 & 21.76 & -2 & 20.08 & 1.19 \\ 
w5347 & ... & 2.405 $\pm$ 0.041 & 0.234 $\pm$ 0.008 & 0.41 & 2.3 & -7.78 & 21.93 & 1.10 & 20.66 & -1 & 19.00 & 1.08 \\ 
w5450 & ... & 2.369 $\pm$ 0.048 & 0.904 $\pm$ 0.005 & 0.79 & 4.0 & -8.78 & 21.09 & 1.25 & 20.29 & -5 & 18.60 & 1.29 \\ 
w5529 & ... & 2.260 $\pm$ 0.055 & 0.349 $\pm$ 0.009 & 0.70 & 2.8 & -8.14 & 22.25 & 1.27 & 21.40 & -5 & 19.72 & 1.22 \\ 
w5577 & ... & 2.484 $\pm$ 0.057 & 0.320 $\pm$ 0.016 & 0.49 & 3.2 & -8.07 & 22.21 & 1.28 & 21.59 & -4 & 19.90 & 1.26 \\ 
w5666 & ... & 2.456 $\pm$ 0.035 & 0.685 $\pm$ 0.005 & 0.67 & 4.0 & -8.45 & 21.34 & 1.29 & 20.74 & -1 & 19.06 & 1.18 \\ 
w5756 & 0.831 & 2.379 $\pm$ 0.026 & 0.577 $\pm$ 0.004 & 0.98 & 4.0 & -8.43 & 21.83 & 1.30 & 21.28 & -5 & 19.59 & 1.29 \\ 
w5840 & ... & 2.405 $\pm$ 0.038 & 0.181 $\pm$ 0.007 & 0.60 & 4.0 & -7.97 & 22.66 & 1.17 & 21.04 & -5 & 19.36 & 1.17 \\ 
w6036 & ... & 2.396 $\pm$ 0.042 & 0.478 $\pm$ 0.012 & 0.82 & 3.6 & -8.30 & 22.01 & 1.38 & 21.22 & -1 & 19.52 & 1.35 \\ 
w6301 & ... & 2.438 $\pm$ 0.059 & 0.438 $\pm$ 0.004 & 0.80 & 4.0 & -8.18 & 21.91 & 1.31 & 21.19 & -5 & 19.50 & 1.23 \\ 
\enddata
\tablenotetext{a}{These are the measured values from fitting a
  elliptical S{\'e}rsic model to the two-dimensional images.}
\tablenotetext{b}{$U-V_z$ refers to the $U-V$ color as if measured by
 $U$ and $V$ filters redshifted to $z=0.831$, the redshift of the cluster.}
\tablenotetext{c}{$T$ refers to the morphological class.  The scheme is
  the same as \citet{postman2005}.  E  $-5 \le T \le -3$ , S0 $-2 \le
  T \le -1$,  with Spirals $0 \le T  \le 10$.  }
\end{deluxetable*}

\clearpage
\newpage

\section{B. Coma Data }

Here we list the compilation of Coma data.  These are galaxies in the
JFK96 catalog with the corresponding selection of  E and S0 galaxies above
a $r$ magnitude limit of $r < 15.1$ mag AB.  Each galaxy has a surface
brightness profile derived from reprocessed SDSS imaging, see
\citet{holden2007}.  This processing mimics what was done for \ms, so
we have comparable data sets.  

As discussed in \citet{holden2007}, each galaxy has a $B_z$ magnitude
derived from the observed $g$ and $g-r$.  Our $B$ values match the
values from JFK96 and we find good agreement between our $B-V$ colors
derived from the $g-r$ imaging and those of \citet{eisenhardt2007}.

The SDSS $u$ imaging is, however, shallow, in comparison to that of
\citet{eisenhardt2007}.  Therefore, we use the colors from that paper.
The colors we list below are extinction corrected and in the
redshifted $(U-V)_z$ filters.  We start with the colors from Table 8 of
\citet{eisenhardt2007}.  We use the extinction corrections from
\citet{schlegel1998}, namely $A_U=0.052$ and $A_V=0.032$.  We redshift
the $U-V$ color with the relation \( (U-V)_z = (U-V) -0.067 (U-V)^2+
0.220 (U-V) -0.220\) as noted in Section \ref{lowz}.

All dispersions come from JFK96.  These were measured in equivalent
manner as the ones at $z=0.83$.

\begin{deluxetable*}{llllllll}
\tablecaption{Data for Galaxies in Coma \label{comadata}}
\tablecolumns{8}
\tablehead{
\colhead{ID} & \colhead{$\log_{10} \sigma$}  &  \colhead{$\log_{10} r_e$} & 
\colhead{$q$} & \colhead{$n$} & \colhead{$-0.4 \mu_B$}  &  \colhead{$B_z$} & \colhead{$(U-V)_z$} \\
\colhead{} & \colhead{$ (\log_{10}{\rm km\ s^{-1}})$ }  &  \colhead{$ (\log_{10}{\rm kpc})$} &
\colhead{} & \colhead{} & \colhead{(-0.4 mag)}  &  \colhead{(mag)} & \colhead{(mag)} \\
}
\startdata
024 & 2.366 & 0.290 & 0.41 & 4.0 & -8.59 & 15.03 & ... \\ 
027 & 2.015 & 0.228 & 0.70 & 2.9 & -8.87 & 16.05 & ... \\ 
031 & 2.416 & 1.230 & 0.58 & 4.0 & -9.57 & 12.79 & ... \\ 
046 & 2.383 & 0.519 & 0.68 & 4.0 & -8.78 & 14.36 & ... \\ 
049 & 2.418 & 0.757 & 0.54 & 4.0 & -8.97 & 13.66 & ... \\ 
057 & 2.232 & 0.431 & 0.65 & 3.4 & -8.71 & 14.63 & ... \\ 
058 & 2.263 & 0.805 & 0.77 & 4.0 & -9.40 & 14.50 & ... \\ 
065 & 2.091 & 0.364 & 0.78 & 2.8 & -9.02 & 15.74 & 1.48 \\ 
067 & 2.215 & 0.014 & 0.91 & 3.2 & -8.44 & 16.04 & 1.44 \\ 
068 & 2.138 & 0.564 & 0.95 & 4.0 & -9.19 & 15.17 & 1.46 \\ 
069 & 2.299 & 0.450 & 0.82 & 3.9 & -8.85 & 14.89 & 1.60 \\ 
070 & 2.199 & 0.313 & 0.96 & 4.0 & -8.78 & 15.40 & 1.48 \\ 
072 & 2.165 & 0.179 & 0.48 & 2.9 & -8.55 & 15.50 & 1.39 \\ 
078 & 2.293 & 0.645 & 0.83 & 4.0 & -9.01 & 14.31 & ... \\ 
081 & 2.183 & 0.456 & 0.72 & 4.0 & -9.16 & 15.64 & ... \\ 
087 & 1.925 & 0.065 & 0.84 & 3.2 & -8.76 & 16.59 & 1.28 \\ 
088 & 2.443 & 0.093 & 0.56 & 4.0 & -8.36 & 15.44 & 1.56 \\ 
098 & 2.200 & 0.332 & 0.88 & 3.6 & -8.79 & 15.33 & ... \\ 
101 & 2.140 & 0.162 & 0.57 & 3.5 & -8.64 & 15.81 & 1.43 \\ 
103 & 2.356 & 0.349 & 0.53 & 4.0 & -8.66 & 14.92 & 1.50 \\ 
104 & 2.301 & 0.195 & 0.37 & 3.6 & -8.58 & 15.50 & 1.51 \\ 
105 & 2.323 & 0.600 & 0.69 & 3.8 & -8.95 & 14.40 & 1.58 \\ 
106 & 2.247 & 0.063 & 0.89 & 3.5 & -8.56 & 16.10 & 1.32 \\ 
107 & 1.852 & 0.256 & 0.82 & 2.1 & -8.93 & 16.06 & 1.30 \\ 
109 & 2.276 & 0.329 & 0.95 & 3.9 & -8.80 & 15.38 & 1.58 \\ 
116 & 2.143 & 0.593 & 0.49 & 4.0 & -9.33 & 15.38 & 1.46 \\ 
118 & 2.237 & 0.478 & 0.87 & 3.5 & -8.92 & 14.92 & 1.51 \\ 
119 & 2.223 & 0.246 & 0.56 & 4.0 & -8.85 & 15.90 & 1.37 \\ 
120 & 2.164 & 0.387 & 0.82 & 3.4 & -8.62 & 14.62 & 1.46 \\ 
121 & 2.340 & 0.077 & 0.70 & 2.2 & -8.54 & 15.97 & 1.47 \\ 
122 & 1.992 & 0.556 & 0.86 & 4.0 & -9.21 & 15.26 & 1.29 \\ 
124 & 2.277 & 0.272 & 0.47 & 2.7 & -8.62 & 15.19 & 1.48 \\ 
125 & 2.267 & -0.061 & 0.70 & 4.0 & -8.33 & 16.14 & 1.35 \\ 
128 & 2.065 & 0.434 & 0.93 & 4.0 & -9.17 & 15.76 & 1.27 \\ 
129 & 2.376 & 1.644 & 0.74 & 4.0 & -10.01 & 11.81 & 1.71 \\ 
130 & 2.352 & 0.489 & 0.58 & 4.0 & -8.88 & 14.78 & 1.50 \\ 
131 & 2.251 & 0.789 & 0.56 & 4.0 & -9.33 & 14.38 & 1.57 \\ 
132 & 2.142 & 0.505 & 0.85 & 4.0 & -9.36 & 15.89 & 1.42 \\ 
133 & 2.375 & 0.264 & 0.62 & 4.0 & -8.55 & 15.08 & 1.46 \\ 
135 & 1.920 & 0.140 & 0.38 & 3.1 & -8.92 & 16.62 & 1.40 \\ 
136 & 2.221 & -0.172 & 0.80 & 2.5 & -8.19 & 16.34 & 1.37 \\ 
137 & 2.245 & 0.364 & 0.62 & 3.6 & -8.71 & 14.96 & ... \\ 
143 & 2.344 & 0.826 & 0.73 & 3.3 & -9.25 & 14.00 & 1.68 \\ 
144 & 2.238 & 0.596 & 0.49 & 4.0 & -9.07 & 14.70 & 1.48 \\ 
145 & 2.152 & 0.415 & 0.39 & 3.3 & -9.01 & 15.47 & 1.48 \\ 
146 & 2.042 & 0.549 & 0.63 & 3.2 & -9.24 & 15.37 & 1.37 \\ 
148 & 2.572 & 1.234 & 0.39 & 4.0 & -9.33 & 12.18 & 1.75 \\ 
150 & 2.044 & 0.353 & 0.42 & 4.0 & -8.94 & 15.61 & 1.40 \\ 
151 & 2.184 & 0.784 & 0.66 & 4.0 & -9.32 & 14.39 & 1.42 \\ 
152 & 2.217 & 0.616 & 0.63 & 4.0 & -9.27 & 15.10 & 1.52 \\ 
153 & 2.153 & 0.359 & 0.89 & 4.0 & -8.97 & 15.65 & 1.44 \\ 
155 & 2.203 & 0.783 & 0.74 & 4.0 & -9.35 & 14.48 & 1.50 \\ 
156 & 2.025 & 0.268 & 0.59 & 3.6 & -9.08 & 16.36 & 1.29 \\ 
157 & 2.142 & 0.305 & 0.54 & 4.0 & -8.97 & 15.92 & 1.32 \\ 
159 & 2.306 & 0.556 & 0.37 & 3.4 & -8.88 & 14.43 & 1.54 \\ 
160 & 2.284 & 0.513 & 0.68 & 3.9 & -9.04 & 15.05 & 1.50 \\ 
161 & 2.251 & 0.547 & 0.53 & 4.0 & -8.95 & 14.65 & ... \\ 
167 & 2.336 & 0.509 & 0.74 & 3.9 & -8.81 & 14.49 & 1.53 \\ 
168 & 2.347 & 0.264 & 0.93 & 2.5 & -8.48 & 14.88 & 1.53 \\ 
170 & 2.174 & 0.445 & 0.52 & 3.7 & -9.00 & 15.30 & 1.44 \\ 
172 & 2.227 & 0.143 & 0.63 & 3.6 & -8.52 & 15.60 & 1.45 \\ 
173 & 2.179 & 0.161 & 0.88 & 4.0 & -8.66 & 15.87 & 1.41 \\ 
174 & 2.287 & 0.014 & 0.24 & 2.3 & -8.28 & 15.64 & 1.50 \\ 
175 & 2.263 & 0.378 & 0.62 & 2.9 & -8.78 & 15.08 & 1.58 \\ 
176 & 2.249 & 0.146 & 0.87 & 1.6 & -8.60 & 15.78 & 1.50 \\ 
177 & 2.038 & 0.218 & 0.18 & 3.7 & -8.93 & 16.25 & 1.40 \\ 
179 & 2.404 & 0.478 & 0.74 & 4.0 & -8.63 & 14.20 & 1.51 \\ 
181 & 2.194 & 0.089 & 0.96 & 3.9 & -8.60 & 16.07 & 1.38 \\ 
191 & 1.994 & 0.007 & 0.87 & 2.3 & -8.54 & 16.32 & ... \\ 
192 & 1.989 & 0.178 & 0.23 & 1.6 & -8.65 & 15.75 & ... \\ 
193 & 2.101 & 0.155 & 0.38 & 2.9 & -8.74 & 16.08 & ... \\ 
194 & 2.398 & 0.646 & 0.68 & 4.0 & -8.93 & 14.10 & ... \\ 
204 & 2.116 & 0.416 & 0.31 & 4.0 & -9.06 & 15.57 & ... \\ 
206 & 2.351 & 0.627 & 0.28 & 3.3 & -8.83 & 13.96 & ... \\ 
207 & 2.192 & 0.181 & 0.24 & 2.6 & -8.67 & 15.79 & ... \\ 
210 & 2.246 & 0.153 & 0.51 & 3.4 & -8.53 & 15.58 & ... \\ 
217 & 2.317 & 0.766 & 0.81 & 4.0 & -9.18 & 14.12 & ... \\ 
238 & 2.041 & 0.037 & 0.86 & 3.4 & -8.51 & 16.10 & ... \\ 
239 & 2.359 & 0.791 & 0.62 & 4.0 & -9.18 & 14.01 & ... \\ 
240 & 2.415 & 1.024 & 0.61 & 4.0 & -9.31 & 13.18 & ... \\ 
\enddata

\end{deluxetable*}

\end{document}